\documentclass[aps,prd,twocolumn,nofootinbib,superscriptaddress]{revtex4-2}

\usepackage{amsmath,amssymb, amsthm,amstext}
\usepackage{natbib}
\usepackage{graphicx}
\usepackage{color}
\usepackage[dvipsnames]{xcolor}
\usepackage{array, enumerate}
\usepackage{bm}
\usepackage{multirow}
\usepackage[breaklinks,colorlinks,citecolor=NavyBlue,urlcolor=NavyBlue, linkcolor=magenta]{hyperref}
\usepackage{braket}
\usepackage{txfonts}
\usepackage[normalem]{ulem}

\def\be{\begin{equation}}
\def\ee{\end{equation}}
\def\ba{\begin{aligned}}
\def\ea{\end{aligned}}

\begin{document}

\title{Common envelope evolution of ultralight Boson Clouds}
\author{Ao Guo}
\email{guoao23@mails.ucas.ac.cn}
\affiliation{International Centre for Theoretical Physics Asia-Pacific, University of Chinese Academy of Sciences, 100190 Beijing, China}
\author{Qi-Yan Zhang}
\email{zhangqiyan22@mails.ucas.ac.cn}
\affiliation{International Centre for Theoretical Physics Asia-Pacific, University of Chinese Academy of Sciences, 100190 Beijing, China}
\author{Huan Yang}
\email{hyangdoa@tsinghua.edu.cn}
\affiliation{Department of Astronomy, Tsinghua University, Beijing 100084, China}
\author{Jun Zhang}
\email{zhangjun@ucas.ac.cn}
\affiliation{International Centre for Theoretical Physics Asia-Pacific, University of Chinese Academy of Sciences, 100190 Beijing, China}
\affiliation{Taiji Laboratory for Gravitational Wave Universe (Beijing/Hangzhou), University of Chinese Academy of Sciences, 100049 Beijing, China}

\begin{abstract}
Ultralight bosons can be excited around spinning black holes via black hole superradiance. These boson clouds may play an important role in the orbital evolution of binary black holes. In this work, we investigate the formation and evolution of common envelopes of ultralight boson clouds in comparable mass-ratio black hole binaries. We describe the cloud evolution using gravitational molecular eigenstates and analyze the possible level transitions during orbital decay, as well as the impact on orbital dynamics. We find that the cloud can generally lead to eccentricity growth. In particular, the eccentricity may vary significantly during level transition, leaving an eccentricity of ${\cal O}(0.1)$ within the detection band of ground-based gravitational wave detectors.
\end{abstract}
\maketitle

\section{Introduction}

Ultralight bosons can arise in many fundamental theories~\cite{Arvanitaki:2009fg} and are proposed as possible dark matter candidates~\cite{Turner:1983he,Press:1989id,Hu:2000ke,Amendola:2005ad,Schive:2014dra,Hui:2016ltb}. If the Compton wavelength of a boson field is comparable to the size of a rapidly rotating black hole, the field can experience superradiant growth, forming a cloud around the black hole~\cite{ZelDovich:1971, Press:1972zz, Zouros:1979iw, Detweiler:1980uk, Brito:2015oca, Arvanitaki:2014wva}. The resulting system, often referred to as a gravitational atom due to its structural similarity to a hydrogen atom, provides a promising laboratory for probing ultralight bosons~\cite{Brito:2017wnc, Brito:2017zvb, Tsukada:2018mbp, Isi:2018pzk, Palomba:2019vxe,Tsukada:2020lgt,Ng:2020ruv,LIGOScientific:2021rnv,Yuan:2022bem, Aurrekoetxea:2023jwk, Aurrekoetxea:2024cqd, Miller:2025yyx}.

\begin{figure}[t]
\includegraphics[width=0.475\textwidth]{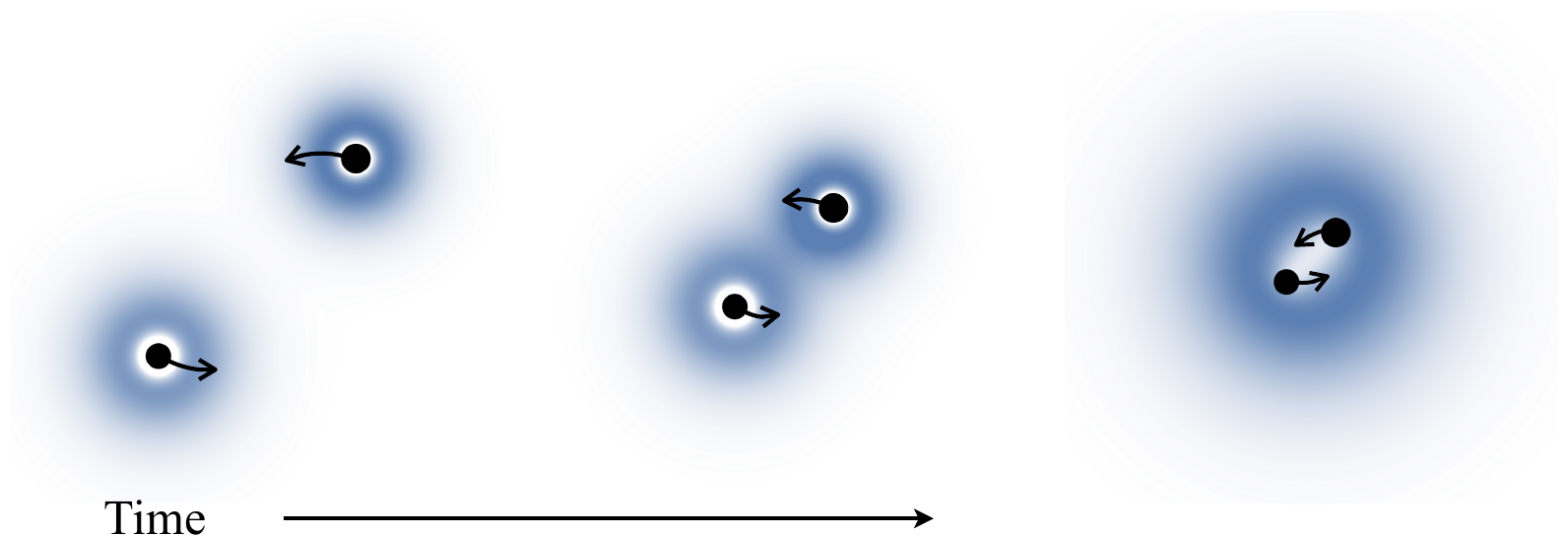}
    \caption{Cloud common envelope formation. For comparable mass-ratio binaries, a cloud is expected to be bounded by its own host black holes when the orbital separation is much larger compared to the cloud. As the orbit decays, the clouds may undergo mass transfer and eventually form a common envelope surrounding the binary.}
\label{fig:cartoon}
\end{figure}

Clouds of ultralight bosons can also form around black holes in a binary system. The evolution of such clouds~\cite{Baumann:2018vus,Baumann:2019ztm,Berti:2019wnn,Ikeda:2020xvt,Choudhary:2020pxy,Tong:2022bbl,Baumann:2021fkf,Guo:2024iye,Liu:2024mzw,Tomaselli:2024ojz} as well as their signatures on orbital dynamics~\cite{Cardoso:2011xi,Ferreira:2017pth,Zhang:2018kib,Zhang:2019eid,Baumann:2022pkl,Tomaselli:2023ysb,Cao:2023fyv, Tomaselli:2024bdd, Cao:2024wby,Takahashi:2024fyq, Zhu:2024bqs,Tomaselli:2024dbw,Boskovic:2024fga,Guo:2023lbv,DeLuca:2021ite,DeLuca:2022xlz,DeLuca:2025bph,Tomaselli:2025jfo} have been extensively studied in the literature. In particular, a cloud in a binary is expected to be bounded by its host black hole, when the orbital separation is much larger compared to the cloud, or if the host black hole is much heavier than the companion black hole; cf. Fig.~\ref{fig:cartoon}. In these cases, the gravitational effects of the companion black hole can be treated perturbatively, and a cloud that is initially in one of the bound states can transit to other bound states~\cite{Baumann:2018vus,Baumann:2019ztm} and even ionization~\cite{Baumann:2021fkf} under the tidal perturbation of the companion black hole. On the other hand, for comparable mass-ratio binaries, it is shown in Refs.~\cite{Guo:2023lbv,Guo:2024iye} that the cloud bounded by one of the binary black holes can start transferring to the region around another black hole, as the orbital separation approaches the size of the cloud. Eventually, one may expect that the cloud (or clouds if both black holes are dressed with their own clouds, which is very likely for comparable mass-ratio binaries) will form a common envelope, surrounding the whole binary. In this case, the system may no longer be treated as a perturbed gravitational atom but is more like a rotating gravitational molecule~\cite{Ikeda:2020xvt}.

In this work, we investigate the formation and evolution of the common envelope state of an ultralight boson cloud in comparable mass-ratio binaries. Focusing on cloud evolution in the late stage, i.e., when the orbital separation is comparable to the cloud size, we describe the cloud evolution with the molecular eigenstates and investigate the possible level transitions between the eigenstates during orbital decay and the potential effects of the cloud on orbital dynamics.

This paper is organized as follows: We first introduce gravitational molecules in Sec.~\ref{sec:GM}, highlighting their relation to the gravitational atoms. Then we set up the framework for analyzing the evolution of gravitational molecules in circular orbits in Sec.~\ref{sec:cecir} and in elliptical orbits in Sec.~\ref{sec:ceecc}. Finally, we discuss the common envelope evolution of superradiant clouds in Sec.~\ref{sec:ce} and its signatures on orbital dynamics in Sec.~\ref{sec:imp}. Section ~\ref{sec:con} devotes to conclusion and discussion. We will take the $(-,+,+,+)$ metric convention and set $\hbar=c=1$. We use $\varphi$ for atomic states and $\psi$ for molecular states.

\section{Gravitational Molecule}\label{sec:GM}

In this section, we shall introduce the gravitational molecule, which is proposed in Ref.~\cite{Ikeda:2020xvt}. We shall consider a real scalar field $\Phi$ of mass $\mu$ in a black hole binary spacetime, the dynamics of which is governed by Klein-Gordon equation
\begin{equation}
    \left(\Box-\mu^2\right) \, \Phi=0. \label{eq:KG}
\end{equation} 
To the leading order of post-Newtonian expansion, the metric of the black hole binary can be written as
\begin{equation}
    {\rm d}s^2=-(1+2\Phi_N)dt^2+(1-2\Phi_N)\left({\rm d}r^2 + r^2 {\rm d} \theta^2 + r^2 \sin^2\theta d\phi^2 \right), \nonumber
\end{equation}  
where
\begin{equation}
    \Phi_N=-\frac{GM_1}{\left|\mathbf{r}-\mathbf{r_1}(t)\right|}-\frac{GM_2}{\left|\mathbf{r}-\mathbf{r_2}(t)\right|}
\end{equation} is the Newtonian potential, and $M_{1,2}$ and $\mathbf{r_{1,2}}$ denote the mass and position of each black hole. In the nonrelativistic limit, we take the ansatz
\begin{equation}\label{eq:psi}
\Phi= \frac{1}{\sqrt{2\mu}}\left(\Psi e^{-i\mu t}+\Psi^*e^{i\mu t}\right) \,,
\end{equation}
where $\Psi$ is a complex field that varies on a timescale much longer than $\mu^{-1}$. Substituting Eq.~\eqref{eq:psi} into Eq.~\eqref{eq:KG} and keeping only the leading terms in the nonrelativistic and weak field limits, the Klein-Gordon equation~\eqref{eq:KG} reduces to a Schr{\"o}dinger equation,
\begin{equation}\label{eq:lab frame}
    i\partial_t \, \Psi\left(t, \mathbf{r} \right)=\left(-\frac{\nabla^2}{2\mu}+\mu \, \Phi_N\right) \Psi\left(t, \mathbf{r} \right) . 
\end{equation}

The Newtonian potential $\Phi_N$ is time dependent. Nevertheless, we can remove the time dependence by working in a corotating frame in the case of circular orbits: We first introduce the lab frame $(t,r,\theta,\phi)$ such that the origin is centered at the center of mass of the binary, and the binary orbit lies in the equatorial plane, and then we define the corotating frame $(t, \bar{r}, \bar{\theta},\bar{\phi})$ with 
\begin{equation}
  \bar{r}=r,\quad \bar{\theta}=\theta,\quad \text{and} \quad \bar{\phi}=\phi-\Omega t,
\end{equation} 
where $\Omega$ is the orbital frequency. In the corotating frame, Eq.~\eqref{eq:lab frame} is rewritten as 
\begin{equation} \label{eq:co frame}
    i\partial_{t} \, \Psi \left(t, \mathbf{\bar{r}} \right) = \left( H_0+H' \right) \Psi \left(t, \mathbf{\bar{r}}\right),
\end{equation}
with
\begin{equation}\label{eq:H0}
  H_0=-\frac{1}{2\mu}\bar{\nabla}^2  -\frac{ GM_1 \mu}{ \left| {\bar{\mathbf{r}}}- {\bar{\mathbf{r}}}_1 \right|}-\frac{GM_2 \mu}{\left| {\bar{\mathbf{r}}}- {\bar{\mathbf{r}}}_2 \right|} \,
\end{equation} 
and
\begin{equation}
H' = i \Omega\, \partial_{\bar{\phi}} \, .
\end{equation} 
As pointed out in Ref.~\cite{Ikeda:2020xvt}, in the absence of $H'$, Eq.~\eqref{eq:co frame} reduces to
\begin{equation} \label{eq:GM}
    i\partial_{t} \, \bar{\Psi} \left(t, \mathbf{\bar{r}} \right) =  H_0 \, \bar{\Psi} \left(t, \mathbf{\bar{r}}\right),
\end{equation}
which resembles the Schr{\"o}dinger equation of the electron in a single electron heteronuclear diatomic molecule. While the eigenstates of Eq.~\eqref{eq:GM} can be obtained numerically for a general mass ratio $q=M_2/M_1$, we will focus on equal-mass binaries ($M_1=M_2=M$ and $q=1$), in most of this work. As we will show later, the asymptotical properties of the eigenstates can be derived analytically for $q=1$, making the equal-mass binaries a good benchmark.

When $q=1$, $\bar{\Psi}$ resembles the wave function of the electron in a hydrogen molecule ion with a ``fine structure constant" $\alpha \equiv GM\mu$ and a ``Bohr radius" $r_B = GM/\alpha^2$. The bound states are known to be discrete in spectrum with each eigenstate labeled by three ``quantum numbers,"
\begin{equation}\label{eq:Psinlm}
\bar{\Psi}_{n \ell m}  \left(t, \mathbf{\bar{r}}\right) = \bar{\psi}_{n \ell m} \left(\mathbf{\bar{r}}\right) e^{-i \epsilon_{n \ell m} t} \, .
\end{equation}

The energy and the wave function of the eigenstates depend on the separation between two black holes $\left| {\bar{\mathbf{r}}}_2- {\bar{\mathbf{r}}}_1 \right| \equiv a$. 
As $a$ approaches zero, the eigenstates approach the eigenstates of the hydrogen atom $\bar{\psi}_{n \ell m}^{(0)}$. It is just that the Bohr radius of the approached hydrogen atom is $r_B/4$, not $r_B$. On the other hand, in the limit of $a \gg r_B$, the clouds should be bounded around the two black holes separately. As a result, the Hilbert space of the gravitational molecule is the direct sum of the two gravitational atoms' Hilbert spaces. In this case, a molecular eigenstate $\bar{\psi}^{(\infty)}_{n \ell m}$ can be approximated by the superposition of two identical atomic eigenstates,
\begin{equation}\label{eq:largea}
   \bar{\psi}^{(\infty)}_{n \ell m} = \bar{\varphi}_{n_s k m} \,  \left(\rho_1,\sigma_1,\chi \right) \pm \bar{\varphi}_{n_s k m}  \, \left(\rho_2,\sigma_2,\chi \right),
\end{equation}
where $\bar{\varphi}_{n_s k m}$ is the eigenstate of a gravitational atom. Note that the wave function $\bar{\varphi}_{n_s k m}$ is obtained in the parabolic coordinates centered on the $i$th black hole, i.e., $\left\{\rho_i, \sigma_i, \chi\right\}$, and is labeled by $n_s$, $k$, and $m$, which are the principal, parabolic and azimuthal quantum numbers, respectively (see Ref.~\cite{Castillo:2008} for details).

\begin{figure}[t]
\includegraphics[height=0.4\textwidth]{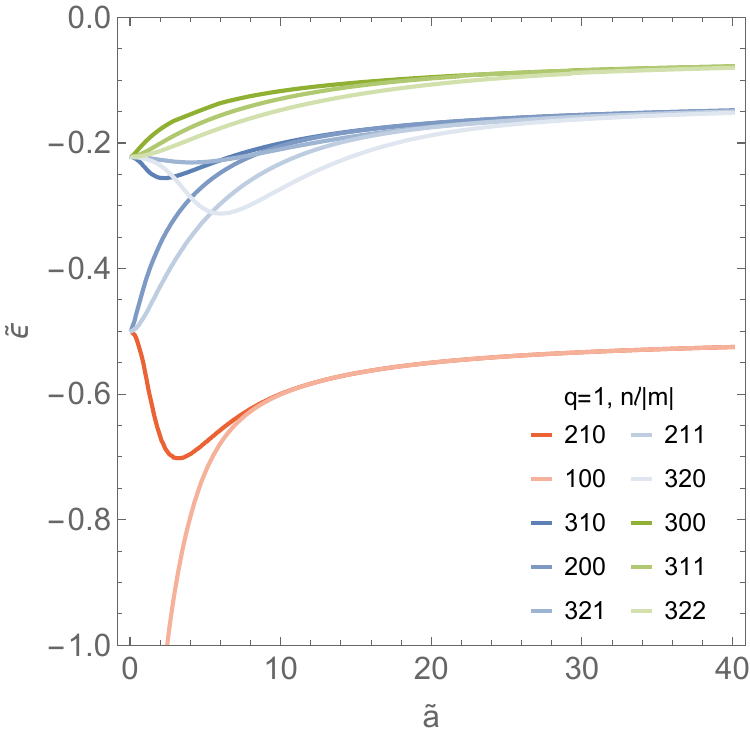}
\includegraphics[height=0.4\textwidth]{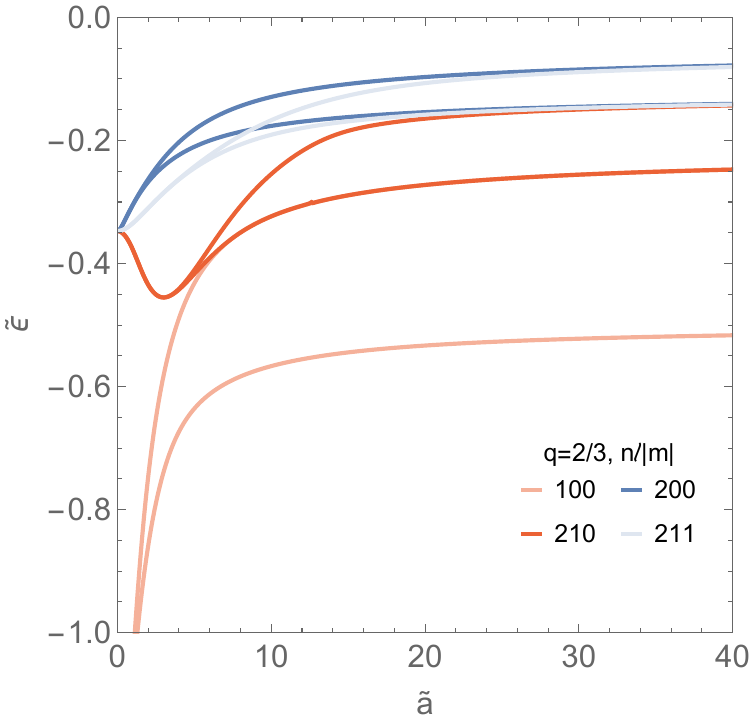}
\caption{Energy levels of the molecular eigenstates. The vertical and horizontal axes show the dimensionless energy $\tilde{\epsilon} \equiv  \epsilon_{n \ell m}/\mu\alpha^2$ and dimensionless orbital semimajor axis $\tilde{a} \equiv a/r_B$. The upper and the lower plots show the energy levels with $q=1$ and $q=2/3$ respectively. For $q=1$, the system is symmetric under reflection, which breaks when $q \neq 1$. As a result, each energy level in the upper plot splits into two levels in the lower plot.}
\label{fig:Enlm}
\end{figure}

Although the molecular eigenstates and the atomic eigenstates in Eq.~\eqref{eq:largea} are labeled with different sets of quantum numbers, they, however, can be related by conserving the number of nodal surfaces of the eigenfunctions~\cite{BATES196813},\footnote{The relation used in this paper is different from the one in Ref.~\cite{BATES196813}. When $|m|$ is even, the superposition of two well-separated atomic states, i.e., $ \bar{\varphi}_{1, n_s k m} \pm  \bar{\varphi}_{2,n_s k m} $ can introduce or eliminate an additional node on the nodal surface where $\zeta=0$, which modifies the relation given by Eqs.(19) and (20) of Ref.~\cite{BATES196813}.}
\begin{equation}
\label{qrelation}
\begin{aligned}
&n_s=n-\frac{l-|m|}{2},  &&k=\frac{l-|m|}{2},  &&\text{if $l-|m|$ is even,}  \\
&n_s=n-\frac{l-|m|+1}{2}, &&k=\frac{l-|m|-1}{2},   &&\text{if $l-|m|$ is odd.} 
\end{aligned}
\end{equation}

The upper plot in Fig.~\ref{fig:Enlm} shows the energy of some molecular eigenstates, where the solid lines correspond to the numerical results obtained with the continued fraction method~\cite{leaver1986solutions} and the shooting method~\cite{grivet2002hydrogen} (see Appendix ~\ref{app:GM} for details).
The numerical results show how the spectrum splits as the orbital separation varies and confirm the relations~\eqref{qrelation} between the quantum numbers. To avoid confusion, we shall label a molecular eigenstate with $n$, $\ell$, and $m$ hereafter, which are also the quantum numbers of the approached atomic eigenstate in the small separation limit. 

In the case of $q \neq 1$, the molecular energy spectrum in the limit of $a \rightarrow \infty$ is still the union of the two energy spectra of two gravitational atoms, but the two atomic energy spectra are no longer identical. The break in the reflection symmetry leads to the energy level split for a given set of ($n$, $\ell$, $m$). To understand the level split, we first recall the case of $q = 1$. In the limit of $a \rightarrow \infty$, the atomic states $\varphi_{1,n\ell m}$ and $\varphi_{2,n\ell m}$ are also molecular eigenstates, and they are degenerate. (Note that they are no longer molecular eigenstates for a finite $a$.) As $a \rightarrow 0$, these two states correspond to the same molecular eigenstate. Now we consider the case of $q\neq 1$. The two atomic states are still molecular eigenstates as $a \rightarrow \infty$, and correspond to the same molecular state as $a \rightarrow 0$, but they are generally not degenerate. Therefore, the energy level denoted by a certain set of ($n$, $\ell$, $m$), the quantum numbers in the small separation limit, should split into two levels and
$a$ deviates from 0. In the lower plot in Fig. \ref{fig:Enlm}, we show the energy levels with mass ratio $q = 2/3$, in which case the reflection symmetry breaks, and each $\epsilon_{n\ell m}$ in the case of $q=1$ splits
into two levels once $a$ deviates from 0.

\section{Cloud Evolution in Circular Orbits}
\label{sec:cecir}

In this section, we investigate cloud evolution in circular orbits with the molecular eigenstates. At a certain orbital separation, the molecular eigenstates form a complete basis of bound states. Therefore, the cloud in the binary can be written as
\be\label{coc}
\bar{\Psi}\left(t, \mathbf{\bar{r}}\right) = \sum_i c_i (t)\,  \bar{\psi}_{i} \left(\mathbf{\bar{r}}\right) e^{-i \epsilon_i t}\, ,
\ee
where $i$ is a short notation for the quantum numbers of the molecular eigenstate.\footnote{In principle, one cloud also includes unbound states, transitions where states could become efficient when $a \sim r_B$. In this paper, we shall focus on interactions between bound states, and we will investigate transitions to unbound states in future study.} In this case, the cloud evolution is fully described by $c_i(t)$. However, in practice, the orbital separation also evolves with time. If the orbit evolution is dominated by gravitational wave radiation, the orbital frequency changes on a timescale 
\be
\tau_{\rm GW} \sim \frac{5}{192} \frac{a^4}{\left(G M\right)^3} \, .
\ee
Comparing to the energy of the eigenstate $\epsilon$, we have $\epsilon \tau_{\rm GW}  \sim  \tfrac{5}{192}\left(\tfrac{a}{GM}\right)^4 \alpha^3$, which is much larger than $1$ if $a/{r_B} \gg (192\alpha^5/5)^{1/4}$. In this case, the orbit decays so slowly that the eigenstates $\bar{\Psi}_i$  evolve adiabatically. In other words, in the absence of $H'$, a cloud superposed by several eigenstates evolves simply by adjusting itself to fit the eigenstates as the orbit decays. This adiabatic condition can be easily satisfied in the large separation limit and could even be satisfied in the small separation limit if $(192\alpha^5/5)^{1/4} \ll 1$, namely $\alpha \ll 0.48$. Therefore, we assume that the adiabatic condition is always satisfied in the following analysis. Moreover, the perturbative treatment of $H'$ requires $H' \ll H_0$, which turns out to be satisfied for the cases studied in this work.

\begin{figure}[t]
\includegraphics[height=0.4\textwidth]{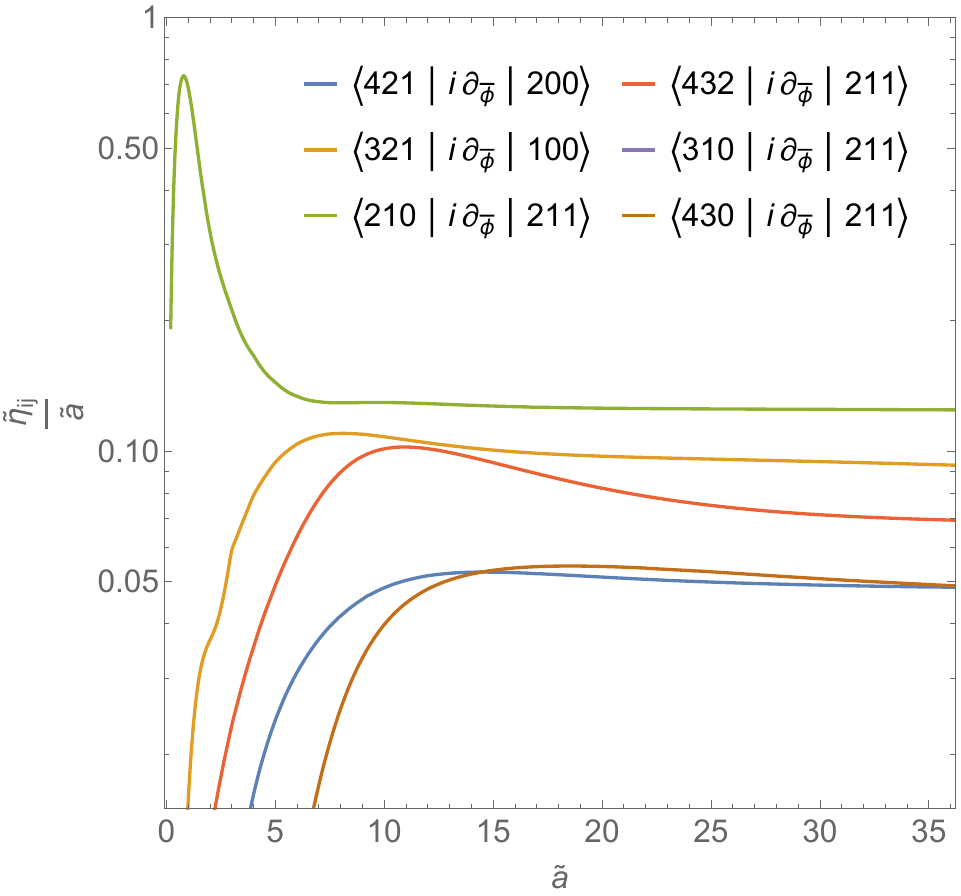}
\caption{Couplings between molecular eigenstates. The vertical axis shows $\tilde{\eta}_{ij}/\tilde{a}$, where $\tilde{\eta}_{ij} = \eta_{ij}/\Omega$ which scales with $\tilde{a}$ at large separation limit.}
\label{fig:eta}
\end{figure}

Now we derive the evolution equation for $c_i(t)$, which can be obtained by substituting Eq.~\eqref{coc} into Eq.~\eqref{eq:co frame},
\begin{equation}
 \label{coceq}
i \dot{c_i} = \sum_j  \eta_{ij} \, e^{i \epsilon_{ij} t }\,  c_j \,,
\end{equation}
where $\epsilon_{ij} \equiv \epsilon_i-\epsilon_j$ is the energy split and $\eta_{ij} \equiv i \Omega \langle \bar{\psi}_i | \partial_{\bar{\phi}}  | \bar{\psi}_j \rangle$ is the coupling strength between different eigenstates. In spheroidal coordinates, we have\footnote{Note there are sign differences comparing to Ref.~\cite{Ikeda:2020xvt}.}
\be\label{eq:dphi}
\ba
\partial_{\bar{\phi}} = \frac{\sqrt{1-\zeta^2}\sqrt{\xi^2-1} \sin \chi}{\zeta^2-\xi^2}\left(\xi \partial_\zeta - \zeta \partial_\xi\right)&  \\
-\frac{\xi \zeta \sqrt{1-\zeta^2}\sqrt{\xi^2-1} \cos \chi}{\left(\xi^2-1\right)\left(1-\zeta^2\right)} \partial_{\chi} & \, ,
\ea
\ee
which indicates $\eta_{ij}$ could be nonzero if
\be\label{selection}
\ba
&m_i - m_j = \pm 1, & \\
&\ell_i +\ell_j= 2p, & \text{for } p\in \mathbb{Z}. 
\ea
\ee
The first rule arises from the periodicity of $\chi$, while the second rule is due to the parity of the eigenstates given $m_i - m_j = \pm 1$. In the small separation limit, the coupling strength approaches
\be
\eta_{ij}^{(0)} = \delta_{n_in_j}\delta_{\ell_i\ell_j}\sum\limits_{m\leq \ell_i}\sum\limits_{m'\leq \ell_j}m \,\delta_{mm'} \,d^{\ell_i*}_{m_im'}\left(\frac{\pi}{2}\right)d^{\ell_j}_{m_jm}\left(\frac{\pi}{2}\right) \, ,
\ee
where $d^{\ell}_{m m'}\left(\frac{\pi}{2}\right)$ is an element of the Wigner D-matrix. In the large separation limit, the wave function is nontrivial only at $\zeta^2 \sim 1- 2 r_B/a$ and $\xi^2 \sim 1+ 2r_B/a$. As a result, $\partial_{\bar{\phi}}$ is dominated by the second term in Eq.~\eqref{eq:dphi}, and thus $\eta_{ij}/\Omega$ scales with $a/r_B$ at a large separation limit. In Fig.~\ref{fig:eta}, we show the coupling strengths obtained with numerical integrations. We find that the coupling strengths approach zero in a small separation limit if the two modes have different $n$ or $\ell$, and all scale with $a$ in a large separation limit, which is consistent with the discussion above.

Estimating the energy difference between the molecular states with $\sim \alpha^3/GM$, one can find $\left|\eta_{ij}/\epsilon_{ij}\right| \sim (r_B/a)^{1/2}$, which means each eigenstate evolves adiabatically when $a \gg r_B$ even in the presence of $H'$. However, nonadiabatic transitions could occur when the energy levels of two coupled states cross each other at a certain orbital separation $a_\times$ as the orbital evolves. This process is known as the Landau-Zener transition.
The transition rate can be estimated as $P_{ij}^{\times} = e^{ -2 \pi Z_{ij}^{\times} }$ with~\cite{Landau1932,1932RSPSA.137..696Z}
\be
Z_{ij}^{\times} \equiv \left. \eta_{ij}^2 \left|\frac{{\rm d} \epsilon_{ij}}{{\rm d} a}\, \dot{a} \right|^{-1} \right|_{a=a_{\times}} \,.
\ee
Assuming the orbit decays by radiating gravitational waves, we have $Z_{ij}^{\times} \propto \alpha^{-5} \gg 1$, which means a complete transition to the destination state for $\alpha \ll 1$. In Fig.~\ref{fig:Enlm}, we show the energy levels with $n \le 3$. We find that the $\ket{320}$ mode intersects with $\ket{21\pm1}$, $\ket{200}$,  and $\ket{310}$, the coupling strengths which, however, are zero due to the selection rules \eqref{selection}. Therefore, $\ket{320}$ does not transit to these modes. We also find that the $\ket{321}$ mode intersects with $\ket{200}$ and $\ket{310}$. Although $\ket{321}$ and $\ket{310}$ do not couple, $\ket{321}$ and $\ket{200}$ cross at $a_{\times} = 8.48 r_B$ with a strength $\eta = 0.35 \alpha^6M^{-2}$, indicating a complete transition to $\ket{321}$.

\section{Cloud Evolution in Eccentric orbits}
\label{sec:ceecc}

In this section, we shall extend the discussion to the case of eccentric orbits. For generic eccentric orbits, we have \begin{equation}
\bar{r}_{1,2} =\frac{a_{1,2}(1-e^2)}{1+e\cos{\phi_*(t)}} \quad \text{with} \quad a_{1,2}=\frac{M_{2,1}}{M_1+M_2}a \, ,
\end{equation}
where $\phi_*(t)$ is the true anomaly of the orbit and $a$ is the semimajor axis that characterizes the orbit separation. For $e \ll 1$, we can expand $H$ in $e$. By doing so, we have
\be
H = H_0 + H' + {\cal O}(e^2) \, .
\ee
While $H_0$ is still given by Eq.~\eqref{eq:H0}, the interaction Hamiltonian becomes
\begin{equation}
\ba
&H'=i \dot{\phi}_* \partial_{\bar{\phi}} - e\cos{u(t)}\\
&\times \left[ \frac{GM_1\mu (a_1^2-a_1\bar{r}\cos{\bar{\phi}})}{|\bar{\mathbf{r}}-\mathbf{a_1}|^{3}} + \frac{GM_2\mu (a_2^2-a_2\bar{r}\cos{\bar{\phi}})}{|\bar{\mathbf{r}}-\mathbf{a_2}|^{3}}\right]  \, ,
\ea
\end{equation}
where $\dot{\phi}_*=\bar{\Omega}t+2e\sin{\bar{\Omega}t}+\mathcal{O}(e^2)$ with 
\be
\bar{\Omega}=\sqrt{\frac{G(M_1+M_2)}{a^3}}
\ee 
being the averaged orbital frequency. Taking ansatz~\eqref{coc}, the evolution of the cloud in eccentric orbits is governed by
\begin{equation}\label{eq:eqce}
    i \dot{c}_i=\sum_{j}\left [\eta_{ij} e^{i\epsilon_{ij}t}+\eta'_{ij}\left(e^{i \epsilon_{ij}t +\bar{\Omega}t}+e^{i\epsilon_{ij}t-\bar{\Omega}t} \right)\right ]c_j,
\end{equation}
where $\eta_{ij}=i\bar{\Omega}\langle \bar{\psi}_i | \partial_{\bar{\phi}}  | \bar{\psi}_j \rangle$ and 
\begin{equation}\label{eq:etae}
\ba
    \eta'_{ij}&=e \eta_{ij}-\frac{\alpha e}{2} \\
   & \times \left\langle \bar{\psi}_i \left| \frac{(a_1^2-a_1\bar{r}\sin{\bar{\theta}}\cos{\bar{\phi}})}{|\bar{\mathbf{r}}-\mathbf{a_1}|^{3}} + \frac{q (a_2^2-a_2\bar{r}\sin{\bar{\theta}}\cos{\bar{\phi}})}{|\bar{\mathbf{r}}-\mathbf{a_2}|^{3}}\right| \bar{\psi}_j \right\rangle \, .
\ea
\end{equation}
For equal-mass binaries, the first term in Eq.~\eqref{eq:etae} is proportional to $\eta_{ij}$, which has been discussed in Sec.~\ref{sec:cecir} with values shown in Fig.~\ref{fig:eta} for some couplings. The second term in Eq.~\eqref{eq:etae}, on the other hand, turns out to be zero when $q=1$. This is because the selection rule with $m_i=m_j\pm1$ requires that the coupled two states have to be of one symmetric and one antisymetric with respect to the plane $\phi=\pm\pi/2$, in which case the gravitational effects of the two black holes cancel in the integral.

\begin{figure*}[t]
\includegraphics[height=0.3\textwidth]{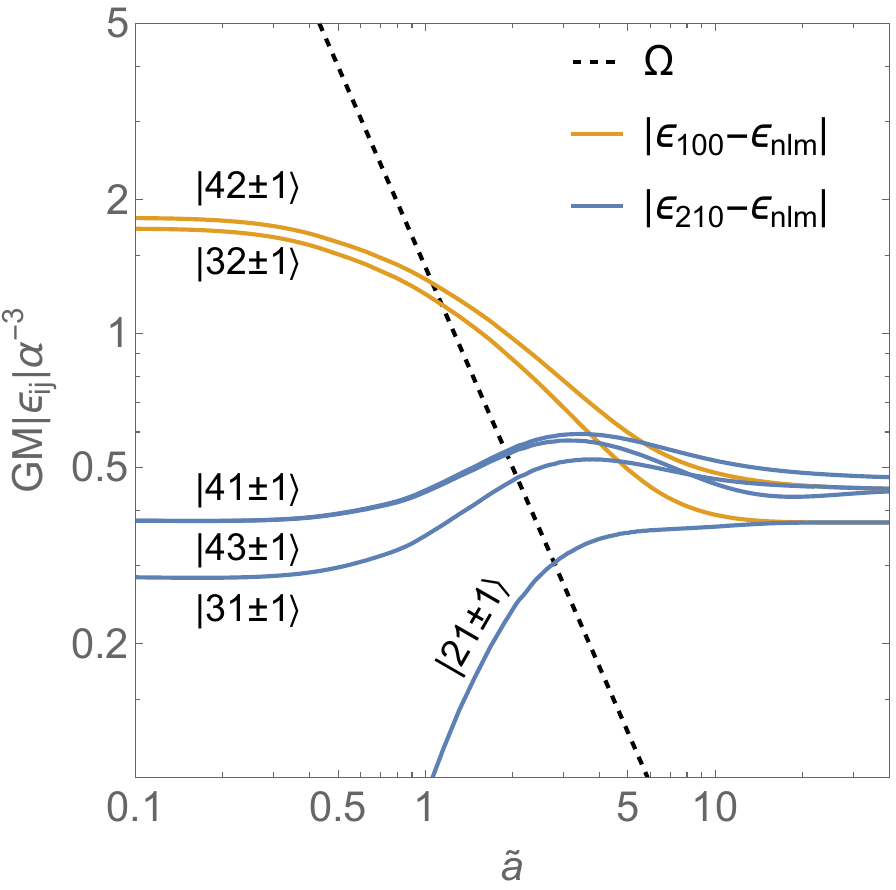}
\includegraphics[height=0.3\textwidth]{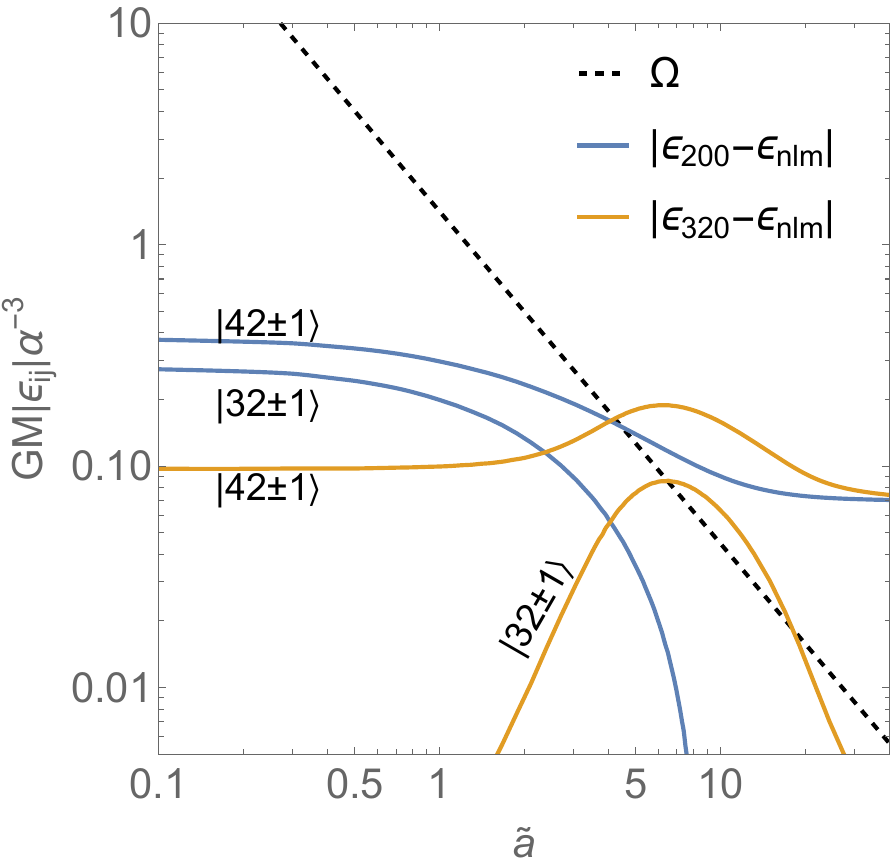}
\includegraphics[height=0.3\textwidth]{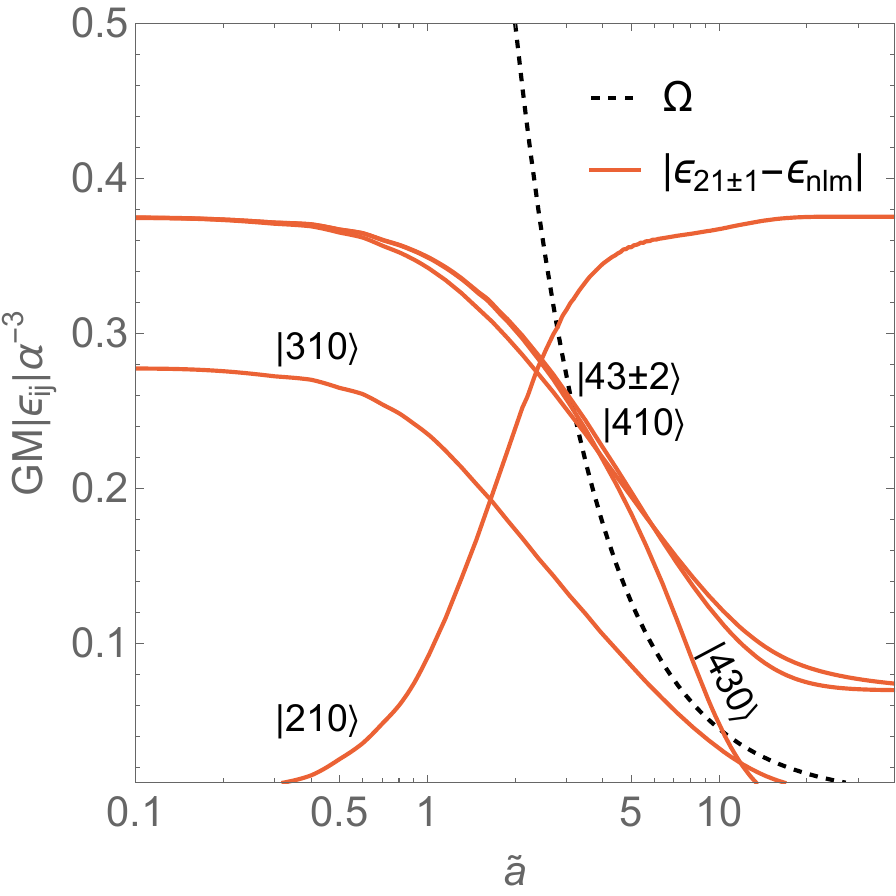}
\caption{Semimajor axis of orbits at eccentricity-induced resonant transitions. In these plots, the colored lines show the energy difference between the considered states ($\ket{100}$, $\ket{210}$, $\ket{200}$, $\ket{320}$, and $\ket{21\pm1}$) and their coupling states with $n \le 4$. The black lines show the averaged orbital frequency. The eccentricity-induced resonant transitions are expected to happen near the intersects of the colored lines and black dashed lines. }
\label{fig:res}
\end{figure*}

According to Eq.~\eqref{eq:eqce}, a nonzero eccentricity could also lead to resonant transitions between two molecular eigenstates. Those transitions occur at $\bar{\Omega} = \pm \epsilon_{ij}$ if $\eta'_{ij} \neq 0$. Near the resonant orbit, that is, orbit with $\bar{\Omega} \simeq \pm \epsilon_{ij}$, one can neglect the highly oscillating terms proportional to $e^{i \epsilon_{ij} t}$ and $e^{i \epsilon_{ij} t \mp i \bar{\Omega}t}$. The transition is then Landau-Zener type, and the transition rate can be estimated by $P_{ij}^{\,e} = e^{ -2 \pi Z_{ij}^{e} }$, where 
\be
Z_{ij}^{e} \equiv \left. {\eta'_{ij}}^{2} \left|\gamma \pm \frac{{\rm d} \epsilon_{ij}}{{\rm d} a}\, \dot{a} \right|^{-1} \right|_{a=a_{e}} \,,
\ee
where $a_e$ is the semimajor axis of the resonant orbit. We show the energy differences between some of the coupled eigenstates and the averaged orbital frequency in Fig.~\ref{fig:res}, from which one can find the orbital separation for the potential transitions. For example, $\ket{210}$ couples to $\ket{n \ell \pm1}$ with odd $\ell$, and could resonantly transit to $\ket{21\pm1}$, $\ket{31\pm1}$, $\ket{43\pm1}$, and $\ket{41\pm1}$ when $a = 2.79 r_B$, $2.08 r_B$, $1.91 r_B$, and $1.89 r_B$, respectively. Given the fact that $a_e \sim r_B$, we have $Z^{e} \propto e^2 \alpha^{-5}$. For example, we find $Z^e \sim 0.24 e^2 \alpha^{-5}$ for the transition to $\ket{21\pm1}$. That is, with $e=0.1$, most of the cloud that is initially in $\ket{210}$ will transit to the states of $\ket{21\pm1}$ in $a \simeq 2.78 r_B$ if $\alpha < 0.37$. Clouds that remain in the $\ket{210}$ state after the resonant transition can transfer to higher-energy states at smaller orbital separations. Similarly, a cloud that is initially in $\ket{100}$ can transit to states of $\ket{n \ell \pm1}$ with even $\ell$.

\section{Common Envelope Evolution of Superradiant Clouds}
\label{sec:ce}

With the framework built in Sec.~\ref{sec:cecir}, we will discuss the evolution of the common envelope phase of superradiant clouds. 

\subsection{Initial conditions}
\label{sec:ceinit}

Superradiant clouds typically grow individually around their own host black holes when the orbit is in a very large separation. In this case, each cloud is mainly in one of its atomic states; therefore, the initial wave function is naturally given by the superposition of the two atomic states. To utilize the framework built in Sec.~\ref{sec:cecir}, we need to decompose such an initial state into molecular eigenstates. To be concrete, we consider two identical black holes inspiral in circular orbit with their spins aligned with the orbital angular momentum. The clouds associated with each black hole are assumed to be dominated by the same atomic eigenstate, and hence the initial wave function of the two clouds can be written as
\begin{equation}
\Psi_0 = e^{-i E_{\rm nlm} t} \left(\varphi_{1,{\rm nlm}} + \varphi_{2,{\rm nlm}}\right) \,,
\end{equation}
where $E_{\rm nlm}$ is the energy of the atomic eigenstate, and $\varphi_{a, \rm nlm}$ is the atomic eigenstate centered on the $a$th black hole. Note that $\varphi_{a, \rm nlm}$ grew via superradiance, and is different from the eigenstate $\bar{\varphi}_{n_s k m}$ in Eq.~\eqref{eq:largea}. Here we use nonitalic quantum numbers to indicate the difference. In particular, ${\rm m}$ characterizes the angular momentum of $\varphi_{a,\rm nlm}$ along the black hole spin, while $m$ characterizes the angular momentum along $\mathbf{r}_2 - \mathbf{r}_1$. To relate the initial state $\Psi_0$ with the corotating molecular eigenstates $\bar{\Psi}_{n\ell m}$, we first decompose $\varphi_{a, \rm nlm}$ into angular momentum eigenfunctions in the direction of $\mathbf{r}_2 - \mathbf{r}_1$, i.e., $\bar{\varphi}_{a, \rm n l m}$. Doing so introduces a time-dependent phase and also mixing between different angular momentum eigenstates,
\begin{equation}
\Psi_0 = e^{- i {\rm m} \Omega_0 t - i E_{\rm nlm} t} \sum_{\rm m'} d_{\rm m m'}^{\rm l}\left(\tfrac{\pi}{2}\right) \left(\bar{\varphi}_{1,{\rm n l m'}} + \bar{\varphi}_{2,{\rm n l m'}}\right) \, ,
\end{equation}
where $d_{\rm m m'}^{\rm l}\left(\tfrac{\pi}{2}\right)$ is an element of the Wigner D-matrix and $\Omega_0$ is the orbital frequency at the start time $t_0$. Then we can decompose the atomic eigenstates in spherical coordinates $\bar{\varphi}_{{\rm n l m'}}$ into the atomic eigenstates in parabolic coordinates $\bar{\varphi}_{n_s k m}$ (see Ref.~\cite{Castillo:2008} for details). Finally, using the relation between the atomic and molecular eigenstates in the large separation limit, we can relabel the molecular eigenstates with $n$, $\ell$ and $m$,
\begin{equation}
\ba
\Psi_0 &=  e^{- i {\rm m} \Omega_0 t - i E_{\rm nlm} t} \sum_{n \ell m}  c_{n\ell m}^\infty\, \bar{\psi}_{n\ell m} \\
& =   \sum_{n\ell m}  c_{n \ell m}^{\infty} \,e^{- i \left(E_{\rm nlm}-\epsilon_i\right) t - i {\rm m} \Omega_0 t} \, \bar{\Psi}_{n\ell m} \,,
\ea
\end{equation}
where $c_{n\ell m}^0$ are constants and the sum of $\bar{\Psi}_{n \ell m}$ only involves a subset of molecular eigenstates. Moreover, in the large separation limit, we have $E_{\rm nlm} \approx \epsilon_i$ (cf.~Fig.~\ref{fig:Enlm}). Thus, we conclude that the initial state can be written as a sum of several corotating molecular eigenstates with a time-dependent phase,
\be
\label{eq:init}
\Psi_0 = e^{-i{\rm m} \Omega t}\, \bar{\Psi}_0 \, .
\ee
For example, if the initial cloud consists of two spin-aligned states $\varphi_{211}$, we have
\be
\varphi_{1,{\rm 211}} + \varphi_{2,{\rm 211}} = \frac{e^{-i \Omega_0 t}}{\sqrt{2}}\left(\bar{\psi}_{211}-\bar{\psi}_{21-1}+\bar{\psi}_{310}-\bar{\psi}_{430}\right) \, .
\ee
Similarly, we have
\be
\ba
\label{eq:init2}
&\varphi_{1,{\rm 211}} - \varphi_{2,{\rm 211}} = \frac{e^{-i \Omega_0 t}}{\sqrt{2}}\left(\bar{\psi}_{32-1}-\bar{\psi}_{321}-\bar{\psi}_{200}+\bar{\psi}_{320}\right)\,, \\
&\varphi_{1,{\rm 21-1}} + \varphi_{2,{\rm 21-1}} = \frac{e^{i \Omega_0 t}}{\sqrt{2}}\left(\bar{\psi}_{211}-\bar{\psi}_{21-1}-\bar{\psi}_{310}+\bar{\psi}_{430}\right)\, ,\\
&\varphi_{1,{\rm 21-1}} - \varphi_{2,{\rm 21-1}} = \frac{e^{i \Omega_0 t}}{\sqrt{2}}\left(\bar{\psi}_{32-1}-\bar{\psi}_{321}+\bar{\psi}_{200}-\bar{\psi}_{320}\right)\, , \\
&\varphi_{1,{\rm 100}} + \varphi_{2,{\rm 100}} =  \sqrt{2} \bar{\psi}_{100}\,, \,\,  \varphi_{1,{\rm 100}} - \varphi_{2,{\rm 100}} = -\sqrt{2} \bar{\psi}_{210} \,,
\ea
\ee
where the eigenstate wave functions are normalized to 1. Here we discuss only the case with two identical gravitational atoms. If the clouds surrounding the two black holes are in different atomic eigenstates or with nonparallel spin alignment, the initial condition can also be constructed in a similar way for equal-mass binaries. Going beyond equal-mass binaries is, however, nontrivial. The initial conditions for $c_i$ in Eq.~\eqref{coc} cannot be obtained analytically. Nevertheless, it can be obtained by projecting the initial wave function on the molecular eigenstates numerically. The initial conditions for clouds in unequal-mass binaries will be left for future investigation.

\subsection{Cloud depletion}

Before forming a common envelope, clouds in binaries may undergo mass transfer. For example, a cloud that is initially in $\varphi_{1,{\rm 211}}$ starts to oscillate between $\varphi_{1,{\rm 211}}$ and $\varphi_{2,{\rm 21-1}}$ when the wave functions of the two states develop a noticeable overlap as the two black holes get closer. As a result, the cloud loses mass to the second black hole via $\varphi_{2,{\rm 21-1}}$. It is claimed in Ref.~\cite{Guo:2023lbv} that, for equal-mass binaries, the cloud depletes rapidly around $a \sim 60 r_B$ due to mass transfer to the $\varphi_{{\rm 21-1}} $ mode of the companion black hole. In this case, the cloud cannot survive to form a common envelope. In the following, we will show that the depletion is not efficient, except for very large $q$.

\begin{figure}[t]
\includegraphics[width=0.4\textwidth]{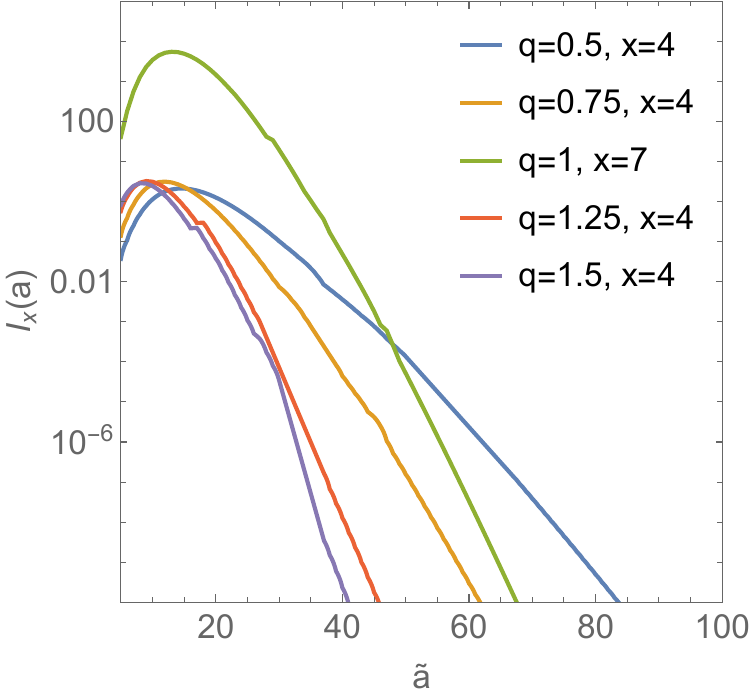}
\caption{Integrals in Eqs. ~\eqref{depconq} and ~\eqref{depcon}. The vertical axis shows  ${\rm I}_x (a) = \left(\frac{a}{r_B}\right)^x G^2M^2 \eta^2/\alpha^6$, where colors of curves represent conditions for various mass ratios ($q=0.5$, $0.75$, $1$, $1.25$ and $1.5$ ) with corresponding powers. }
\label{fig:AA}
\end{figure}

The calculation is based on the formula in ~\cite{Guo:2024iye}: focusing on the transfer between $\varphi_{1,{\rm 211}}$ and $\varphi_{2,{\rm 21-1}}$, the cloud can be described by
\be
\varphi =  c_1 (t) \, \varphi_{1,{\rm 211}} + c_2(t) \,  \varphi_{2,{\rm 21-1}} \,.
\ee
The cloud loses mass to the second black hole at a rate of
\be
- 2 \Gamma_{2, {\rm 21-1}} |c_2|^2 M_c
\ee
During a nonresonant evolution, we have
\be
|c_2|^2 \approx \frac{\eta^2}{\omega^2 + \eta^2} \sin^2\left( \int_{-\infty}^t dt' \sqrt{\omega^2 + \eta^2} t'\right) \, .
\ee
where $\eta$ is the coupling strength, $\omega \approx \Delta \epsilon + 2\Omega$ with $\Delta \epsilon$ being the energy split between $\varphi_{1,{\rm 211}}$ and $\varphi_{2,{\rm 21-1}}$, and the integral over $t'$ takes into account the slowly changing of $\omega$ and $\eta$. 
For $q \neq 1$, we have $|\Delta \epsilon| \approx (1-q^2)\alpha^3/8  \gg |2 \Omega|$, and the clouds deplete significantly only if
\be\label{depconq}
 \Gamma_{2, {\rm 21-1}} \frac{ \eta^2 }{\omega^2} \frac{\Omega}{\dot{\Omega}} \approx \frac{40}{3} \frac{(1+q)^{1/3} q^8}{(1-q^2)^2}  \alpha^2 {\rm I}_4 (a)  \gtrsim 1,\, 
\ee
where we have averaged $|c_2|^2$ over a timescale much longer than $\sqrt{\omega^2 + \eta^2}$ but shorter than $\Omega/\dot{\Omega}$. We also defined ${\rm I}_4 (a) = \left(\frac{a}{r_B}\right)^4 G^2M^2 \eta^2/\alpha^6$ which depends only on $q$. Figure ~\ref{fig:AA} shows the numerical value of ${\rm I}_4 (a)$ for different $q$, from which we conclude that the clouds barely deplete unless $q$ is relatively large. For example, by requiring the maximum of the left-hand side of the inequality~\eqref{depconq} to be less than $1$, we get $q > 3.27$ for $\alpha = 0.1$. The case of $q=1$ is especially, in the sense that $|\Delta \epsilon| \simeq 0$, and $\omega$ is dominated by $2\Omega$.\footnote{To be precise, $|\Delta \epsilon  | \propto \mu \alpha^5 $ and depends on the black hole spin.} In this case, the depletion condition becomes
\be\label{depcon}
 \Gamma_{2, {\rm 21-1}} \frac{ \eta^2 }{\omega^2} \frac{\Omega}{\dot{\Omega}} \approx \frac{5}{96} 2^{1/3}  \alpha^2 {\rm I}_7 (a)  \gtrsim 1,
\ee
where ${\rm I}_7 (a) = \left(\frac{a}{r_B}\right)^7 G^2M^2 \eta^2/\alpha^6$. As shown in Fig.~\ref{fig:AA}, even for $q=1$, the clouds cannot completely deplete. Compared to the estimation in Ref.~\cite{Guo:2023lbv}, $\omega$ differs by $2\Omega$, which is due to binary rotation that is not considered in Ref.~\cite{Guo:2023lbv}. Moreover, the mass transfer can be resonant at certain orbits as discussed in Ref.~\cite{Guo:2024iye}, which, however, do not deplete the cloud completely either. Therefore, we conclude that, for the general mass ratio, the clouds generically survive to form common envelopes in the late inspiral.

\section{Implications on orbital evolution}
\label{sec:imp}

In this section, we discuss the effects of clouds on orbital evolution. Assuming that the cloud in a binary can be described as the superposition of several molecular eigenstates, cf.~Eq.~\eqref{coc}, the binding energy of the cloud can be written as
\begin{equation}
\begin{aligned}
        E_{\rm c}=M_{\rm c}\alpha^2 \left[\sum_{i}|c_i|^2\tilde{\epsilon_i} +  (1+q)^{1/2}\tilde{a}^{-3/2} \sum_{i, j}  c_i^* c_j   \tilde{\eta}_{ij} \right],
        \label{ECLOUD}
\end{aligned}
\end{equation}
where $M_c$ is the cloud mass, and $\tilde{\epsilon}_i$ and $\tilde{\eta}$ are the dimensionless molecular eigenenergy and coupling strength, for example, shown in Fig.~\ref{fig:Enlm} and Fig.~\ref{fig:eta}, respectively. The cloud also contributes to the total angular momentum with 
\begin{equation}
\begin{aligned}
    J_{\rm c} =  -  \frac{G M M_c}{\alpha} \sum_{i, j}  c_i^* c_j   \tilde{\eta}_{ji} \,
\end{aligned}
\label{EJ}
\end{equation}
in the direction that is perpendicular to the orbital plane. The molecular states are the eigenstates of the angular momentum along $\bar{\bf r}_2- \bar{\bf r}_1 $, and hence have zero angular momentum in the direction perpendicular to the orbital plane. In other words, a cloud should typically be a superposition of at least two molecular states.

\begin{figure}
    \centering
    \includegraphics[width=0.4\textwidth]{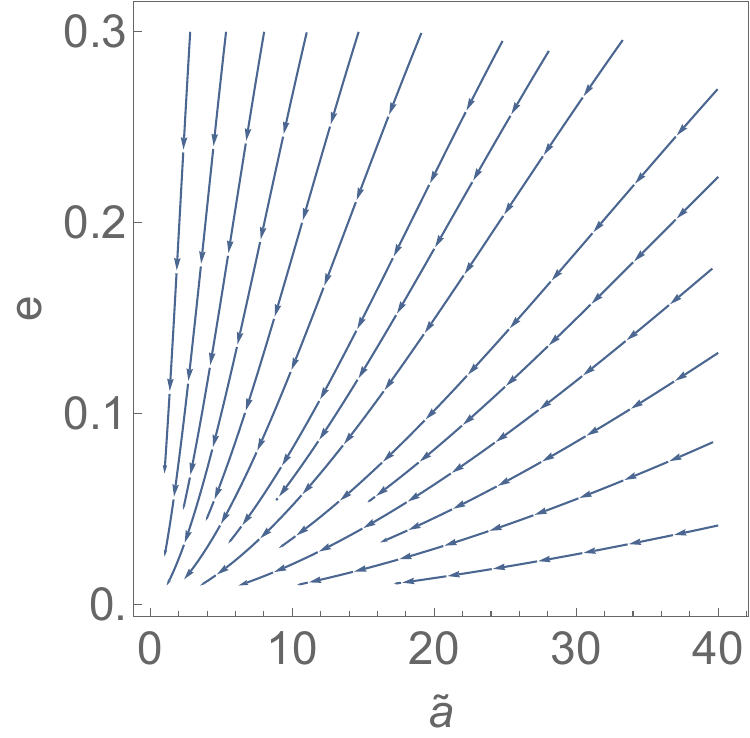} \\
     \includegraphics[width=0.4\textwidth]{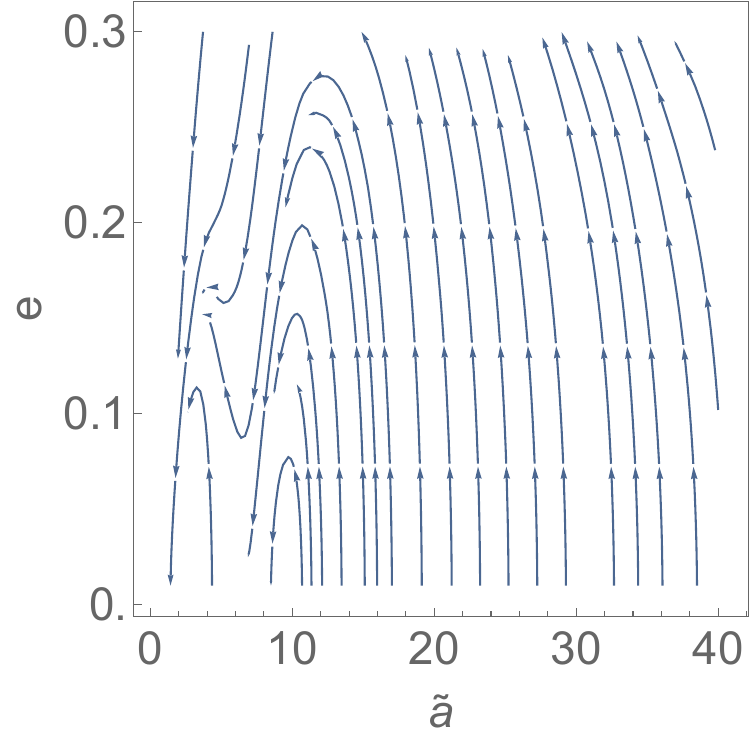}
     \includegraphics[width=0.4\textwidth]{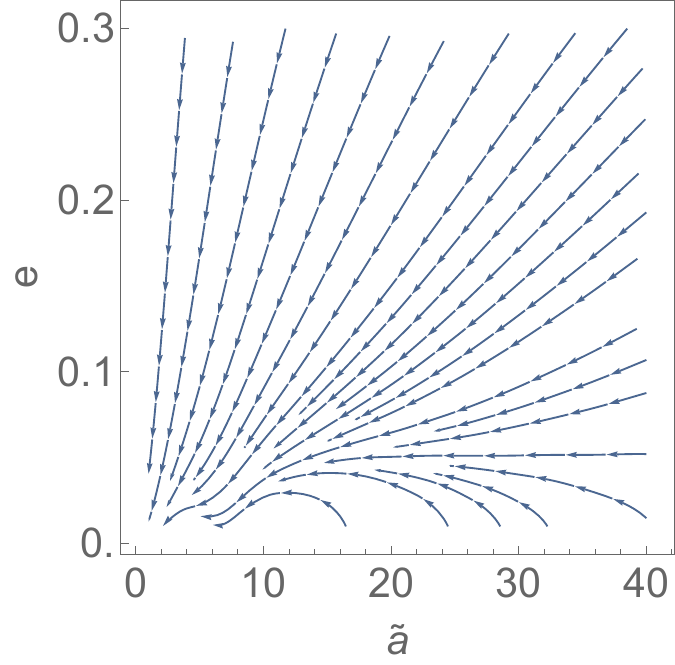}
    \caption{Orbital evolution in the phase space. The upper plot shows the evolution trajectories in the absence of clouds, where the eccentricity always decreases during the orbital decay. The middle and lower plots show the evolution trajectories in the presence of the cloud, where we assume a cloud of $M_c = 0.1 M$ and $0.01 M$, respectively, originating from two spin-aligned superradiance states $\varphi_{1, 211}$ and $\varphi_{2, 211}$, and undergoing adiabatic evolution. In these cases, the eccentricity may increase as the orbits decay.}
    \label{fig:ea}
\end{figure}

We assume that the orbit decays by radiating gravitational waves. Assuming quadrupole radiation, the orbital parameters evolve as
\be
     \frac{d\tilde{a}}{d\tilde{t}}=-\frac{64 q (1+q)f(e) }{5\tilde{a}^3(1-e^2)^{7/2} (1+B)}\,,    
     \label{dadt} 
\ee
\be
\frac{d e}{d\tilde{t}}=  \frac{-304q(1+q) }{15\tilde{a}^4 (1-e^2)^{5/2}}\left[g(e) e +\frac{6f(e)}{19 e}\frac{A-\sqrt{1-e^2}B}{\sqrt{1-e^2}(1+B)}\right]  
 \label{dedt} 
\ee
with
\begin{equation}
    \begin{aligned}
        &B=  \frac{\tilde{M}_c}{q}  \left[ \sqrt{(1+q)\tilde{a}}\left(3 \frac{\tilde{J}_c}{\tilde{a }} -2  \frac{d\tilde{J}_c}{d\tilde{a}}   \right) + 2 \frac{d\tilde{\epsilon}}{d\tilde{a}} \tilde{a}^2 \right]\,, \\
    &A= \frac{\tilde{M}_c}{ q}\sqrt{(1+q)\tilde{a}}\frac{d\tilde{J}_{\rm c}}{d\tilde{a}} \, ,\\
    &f(e)= 1+\frac{73}{24}e^2+\frac{37}{96}e^4 \,
  \quad {\rm and} \quad g(e) =  1+\frac{121}{304}e^2 \,,
    \end{aligned}
    \label{para}
\end{equation}
where we have defined $\tilde{t} \equiv  \alpha^8 t/GM$ and $\tilde{M}_c \equiv M_c/M$.
According to Eq.~\eqref{dadt}, the cloud may decelerate or accelerate the orbit decay, depending on the sign of $B$. Although in principle both $c_i$ and $\tilde{\epsilon}_i$ vary with $a$ as the orbit decays, we expect that $c_i$ does not evolve during adiabatic evolution. In this case, the cloud tends to accelerate orbit decay when $a > 6 r_B$. When $a < 6 r_B$, some of the molecular states tend to decelerate orbit decay. Given the fact $M_c \ll M$, we expect $|B| < 1$, and hence no outspiral caused by the cloud.

According to Eq.~\eqref{dedt}, the cloud plays an important role in the eccentricity evolution during orbital decay. In particular, the right-hand side of Eq.~\eqref{dedt} can be positive in the presence of the cloud, resulting in increased orbital eccentricity. For example, we consider the cloud that initially consists of two spin-aligned atomic states $\varphi_{1, 211}$ and $\varphi_{2, 211}$. During adiabatic evolution, we expect that the coefficients $c_i$ are constant in time and that the orbital evolution can be obtained by solving Eqs.~\eqref{dadt} and ~\eqref{dedt}. In Fig.~\ref{fig:ea}, we plot the evolution trajectories, assuming $\tilde{M}_c = 0.1$. We find that eccentricity may increase when $4 < \tilde{a}<6 $ and $10 < \tilde{a} $.

\begin{figure}
    \centering
    \includegraphics[width=0.4\textwidth]{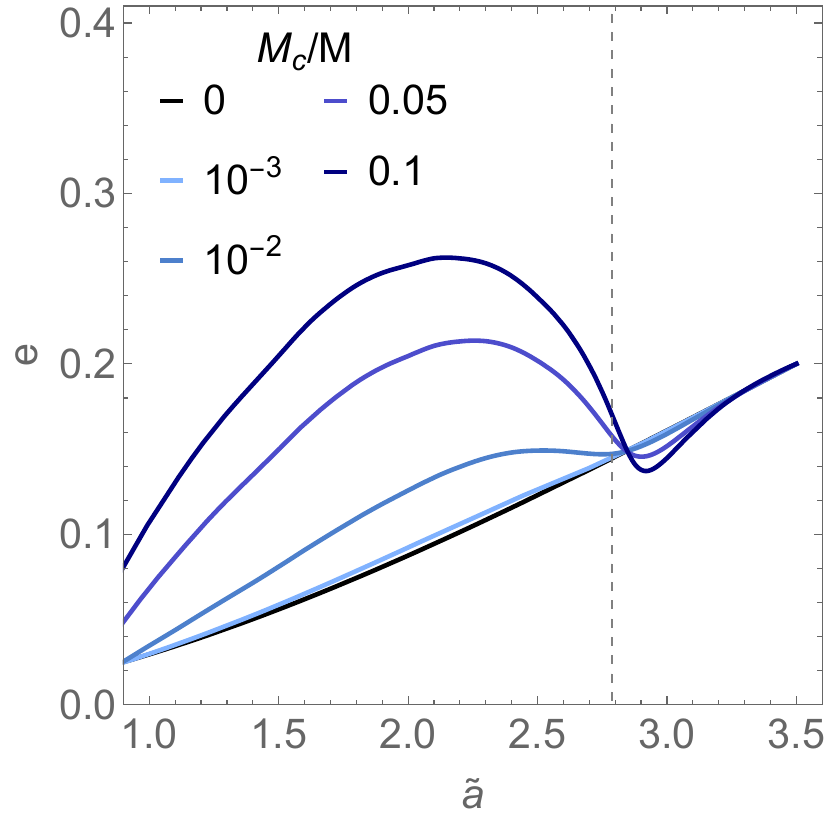} \\
     \includegraphics[width=0.4\textwidth]{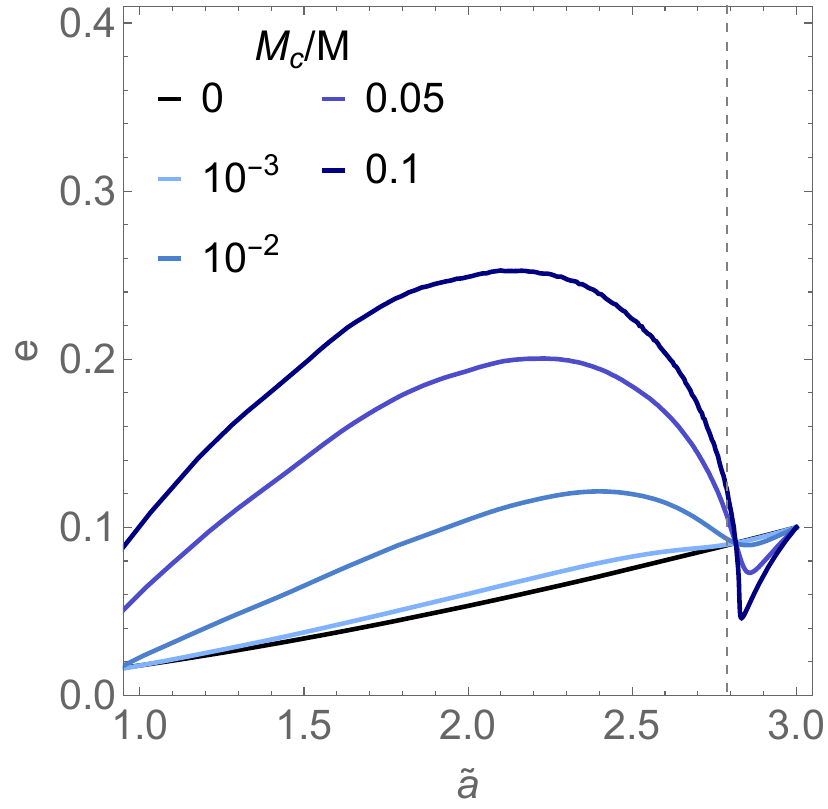}
    \caption{Eccentricity pumping during molecular level transitions. The plots show orbital evolution near the molecular level transition from $\bar{\psi}_{211}$ to $\bar{\psi}_{210}$, assuming different cloud masses. The upper plot assumes an initial eccentricity of $0.2$ at $\tilde{a} = 3$, while the lower plot assumes an initial eccentricity of $0.1$ at $\tilde{a} = 3$. The gray dashed line located at $a\sim 2.79 r_B $ represents the resonant radius of the transition. The evolution is obtained by solving Eqs.~\eqref{dadt}, ~\eqref{dedt}, and ~\eqref{para} numerically. }
    \label{fig:evolution}
\end{figure}

The growing eccentricity may also occur during nonadiabatic evolution, where $c_i$ change significantly due to resonant transitions of the cloud. For the cloud considered above, there is a resonant transition from $\bar{\psi}_{211}$ to $\bar{\psi}_{210}$ at $\tilde{a} \simeq 2.79$. In Fig.~\ref{fig:evolution}, we show the eccentricity evolution near the resonant orbit by solving Eqs.~\eqref{dadt}, ~\eqref{dedt}, and ~\eqref{para} numerically. We find that for $\tilde{M}_c > 10^{-2}$, the eccentricity varies significantly during the transition, leaving an eccentricity of ${\cal O}(0.1)$ at $a \sim r_B$.

Assuming equal-mass binaries, the frequency of the gravitational wave emitted at $a \sim r_B$ can be estimated as
\be
f_B \sim 18 \times \left(\frac{\alpha}{0.1}\right)^3 \left(\frac{5 M_\odot}{M}\right) {\rm Hz}\,.
\ee
In other words, at $f_{B} \sim 20 {\rm Hz}$, the eccentricity of the binary can be estimated as (assuming $\sim f^{-19/18}$ decay from GW radiation from the peak \cite{Peters1964a} )
\begin{align}
    e \sim \mathcal{O}(0.1) \times \left(\frac{\alpha}{0.1}\right)^3 \left(\frac{5 M_\odot}{M}\right)\,.
\end{align}
Therefore, for stellar mass binaries, the superradiant cloud can give rise to a nonzero eccentricity detectable in the LIGO-Virgo-KAGRA (LVK) band. Interestingly, recent analysis of the LVK event GW200105 indicates a median orbital eccentricity of $e \sim 0.145$ at an orbital period of $0.1$s \cite{Morras:2025xfu}. Our work provides a potential origin of the median eccentricity.\footnote{Note that, in practice, the mass of the boson field is a constant, and $\alpha$ varies with black hole masses. In other words, one has $f_B \propto \mu^3 M^2$. Nevertheless, we do not expect to have eccentricity excitation in binaries with heavier black holes, because the superradiant growth is suppressed or the cloud might have already depleted. Therefore, the eccentricity excitation mechanism could explain the observed eccentricity in GW200105, while being in consistence with other LVK events.}

 Note that the eccentricity excitation mechanisms discussed above are different from the one in Refs.~\cite{Tomaselli:2024dbw,Boskovic:2024fga}, which is caused by the level transitions between atomic states and happens only in an inclined orbit and typically at a relatively large orbital separation. As a result, the induced eccentricity hardly survives to the LVK band and cannot be detected with the current ground-based gravitational wave detectors.

\section{Conclusion and Discussion}
\label{sec:con}

In this work, we investigate the evolution of the cloud in the stage where the orbital separation becomes comparable to the cloud size. We describe the cloud evolution with the gravitational molecular eigenstates. With this framework, we discuss the common envelope formation of the superradiant clouds and their late evolution in binaries. We find that most of the time the cloud evolves adiabatically; i.e., each eigenstate slowly adjusts itself as the orbit decays. Resonant transitions between different eigenstates may also happen when the energy levels cross or if the orbital frequency matches the energy difference between the eigenstates. 

We have also investigated the potential effects on the orbital evolution of the clouds. We find that for elliptical orbits, the eccentricity may increase when the orbital separation is greater than 5 times the Bohr radius. This result proposes a potential origin for the median eccentricity at a small orbital separation that was recently implied by the LVK event GW200105~\cite{Morras:2025xfu}. The cloud-induced eccentricity could be further tested with future gravitational-wave observation, providing an alternative way of searching for the ultralight boson field with gravitational wave observations. 

This work provides a framework for investigating cloud evolution and its effects on binary black hole mergers as environments. While in this work we focus on the evolution of bound states, the bound states could also couple to unbounded states through the perturbation $H'$, thereby exciting bosonic radiation. This leads to cloud ionization and extra dissipation of the binary binding energy, which can also be treated within the present framework. Cloud ionization in comparable mass black hole binaries has recently been observed in numerical simulations and analyzed in Ref.~\cite{Guo:2025pea}. In particular, Ref.~\cite{Guo:2025pea} shows that ionization occur in eccentric orbits. In principle, cloud ionization may influence orbital evolution and should eventually be taken into account in a complete treatment. However, ionization becomes efficient only when the orbital separation approaches the Bohr radius, i.e., $\tilde{a} \sim 1$. By contrast, the eccentricity growth identified in this work could occurs at much larger separations, e.g., $\tilde{a} > 20$ (cf. Fig.~\ref{fig:ea}), well before ionization becomes relevant. We therefore do not expect ionization to significantly modify our results in the large-separation regime. At small orbital separations, however, cloud ionization should be properly accounted for. It may deplete the cloud efficiently and quantitatively change the strength of resonant eccentricity pumping (cf.~Fig.~\ref{fig:evolution}). A detailed study of ionization in gravitational molecules will be presented in future work. Together with ionization, the cloud environmental effects on gravitational waves from binary black hole mergers are promising directions for future study.

\section*{ACKNOWLEDGMENTS}

The authors thank Richard Brito, Yifan Chen, and Giovanni Maria Tomaselli for useful discussions. J. Z. is supported by the scientific research starting grants from the University of Chinese Academy of Sciences (Grant No.~118900M061), the Fundamental Research Funds for the Central Universities (Grants No.~E2EG6602X2 and No.~E2ET0209X2), and the National Natural Science Foundation of China (NSFC) under Grant No.~12147103.

\section*{DATA AVAILABILITY}
The data that support the findings of this article are openly available ~\cite{Guodata:2026}.

\appendix

\section{EIGENSTATES OF GRAVITATIONAL MOLECULES}
\label{app:GM}

To find the eigenstates of $H_0$, we introduce the prolate spheroidal coordinates \{$\zeta,\ \xi,\ \chi$\} 
\begin{equation}
    \begin{aligned}
            \bar{x}&=\frac{a}{2}\zeta\xi, \\
            \bar{y}&=\frac{a}{2}\sqrt{1-\zeta^2}\sqrt{\xi^2-1}\sin{\chi},\\
            \bar{z}&=\frac{a}{2}\sqrt{1-\zeta^2}\sqrt{\xi^2-1}\cos{\chi},
            \end{aligned}
\end{equation}
with $a$ being the semimajor axis. The Laplacian in this coordinate can be written as
\begin{equation}
\begin{aligned}
\nabla^{2} =& \frac{4}{a^2(\xi^2 - \zeta^2)} \Bigg\{ \frac{\partial}{\partial \xi} \Big[ (\xi^2 - 1) \frac{\partial}{\partial \xi} \Big] + \frac{\partial}{\partial \zeta} \Big[ (1 - \zeta^2) \frac{\partial}{\partial \zeta} \Big] \\
&+ \Bigg[ \frac{1}{\xi^2 - 1} + \frac{1}{1 - \zeta^2} \Bigg] \frac{\partial^2}{\partial \chi^2} \Bigg\}.
\end{aligned}
\end{equation}
Taking the ansatz
\begin{equation}
    \bar{\psi}_{n\ell m}\left(\zeta,\ \xi,\  \chi\right)=e^{i m \chi}R_{n\ell}(\xi)S_{\ell m}(\zeta),
\end{equation}
for $\bar{\psi}_{n\ell m}$ in Eqs.~\eqref{eq:Psinlm},~\eqref{eq:GM} becomes separable and leads to
\begin{equation}
    \begin{aligned}
        \partial_\zeta\left[(1-\zeta^2)\partial_\zeta S\right]+\left(A_{\ell m}-\frac{\tilde{a}^2\tilde{\epsilon}}{2}\zeta^2+\tilde{a}(1-q)\zeta-\frac{m^2}{1-\zeta^2}\right)S&=0,\\
        \partial_\xi\left[(\xi^2-1)\partial_\xi R\right]+\left(-A_{\ell m}+\frac{\tilde{a}^2\tilde{\epsilon}}{2}\xi^2+\tilde{a}(1+q)\xi-\frac{m^2}{\xi^2-1}\right)R&=0,
        \label{GA2}
    \end{aligned}
\end{equation}
where $A_{\ell m}$ is the angular separation constant. For $q =1$, the system is symmetric under reflection, i.e., $\zeta \rightarrow -\zeta$, while for $q \neq 1$ the symmetry breaks, and the system is covariant under $\zeta \rightarrow -\zeta$ and $q \rightarrow 1/q$.

\subsection{Equal-mass system}

\begin{figure}
    \centering
    \includegraphics[width=0.8\linewidth]{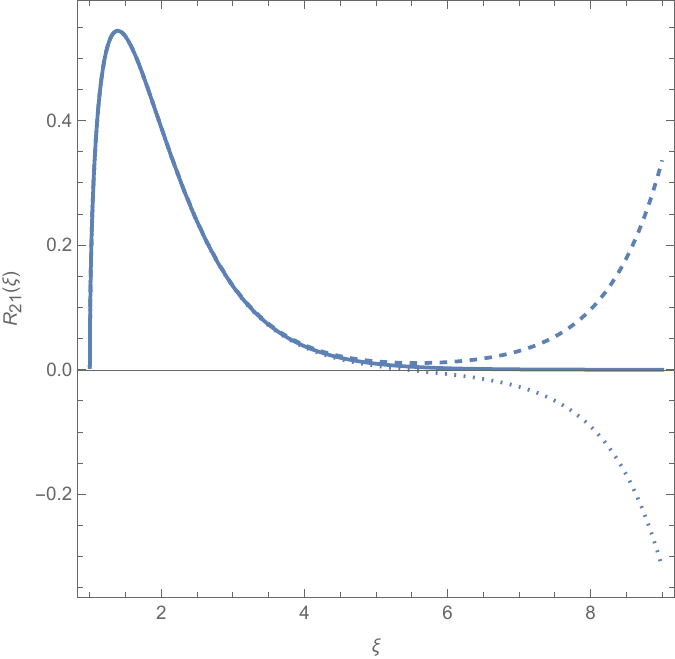}
    \caption{The shooting method used to determine the eigenenergy of the gravitational molecule. The vertical axis is the value of the radial part $R_{21}(\xi)$ of the molecular state $\bar\Psi_{21\pm1}$. The dashed, solid, and dotted lines correspond to $\tilde{\epsilon}=-0.702149$, $-0.701638$, and $-0.70119$, and the middle one is the best value for its convergence. }
    \label{fig:Shoot}
\end{figure}

For equal-mass binaries, the angular part of $\psi_{n\ell m}$, $S_{\ell m}(\zeta)$, turns out to be the spheroidal harmonics. To obtain the radial function $R_{n\ell}(\xi)$, we first introduce
\begin{equation}
   \tilde{R}(\xi) =  (\xi^2-1)^{-|m|/2} R(\xi).
\end{equation}
In the limit of $\xi \rightarrow 1$, Eq.~\eqref{GA2} leads to
\begin{equation}\label{bc1}
  \partial_\xi \tilde{R} =-\frac{2\tilde{a}(1+q)+\tilde{a}^2\tilde{\epsilon}/2+|m|(|m|+1)-A_{\ell m}}{2(|m|+1)}\tilde{R}.
\end{equation}
As $\xi \rightarrow \infty$, we expect $\tilde{R}$ approaches as we look for bound states. Then the radial function $R$ and the eigenenergy of the molecular states can be obtained by solving Eq.~\eqref{GA2} with the shooting method: We integrate Eq.~\eqref{GA2} starting from $\xi_i$, which is very close to $1$, with the boundary condition derived from~\eqref{bc1}. $\tilde{R}(\xi_i)$ could be chosen as any constant, and will be fixed later with the normalization condition
\begin{equation}
     \langle \bar{\psi}_{n\ell m}|\bar{\psi}_{n'\ell' m'}\rangle=\delta_{nn'}\delta_{\ell \ell'}\delta_{mm'}.
\end{equation}
Then by fine-tuning $\tilde{\epsilon}$, we can find a solution that approaches $0$ at large $\xi$ (see Fig.~\ref{fig:Shoot} for illustration). Here $A_{\ell m}$ is determined by $\tilde{\epsilon}$ given the spheroidal harmonics.

\subsection{General mass-ratio system }
\label{app:CF}
For $q\neq1$, the eigenvalues are obtained with the continued fraction method. Following Ref.~\cite{leaver1986solutions}, we introduce
\begin{equation}
\begin{aligned}
        &\kappa=\zeta+1,\quad F(\kappa)=\left(-\kappa^2+2\kappa\right)^{-m/2}S(\kappa-1),\\
        &\lambda=\xi+1,\quad G(\lambda)=\left(\lambda^2-2\lambda\right)^{-m/2}R(\lambda-1) \, ,
        \label{kappa}
\end{aligned}
\end{equation}
and expand $F(\kappa)$ and $G(\lambda)$ as
\begin{equation}
\begin{aligned}
            F(\kappa)=&e^{-\sqrt{-2\tilde{a}^2\tilde{\epsilon}}\kappa}\sum_{n=0}^{\infty}C^{\kappa}_n\kappa^n,\\
            G(\lambda)=&e^{-\sqrt{-2\tilde{a}^2\tilde{\epsilon}}\lambda}\lambda^{-1-m-\frac{a(1-q)}{\sqrt{-2\tilde{a}^2\tilde{\epsilon}}}}\sum_{n=0}^{\infty}C^{\xi}_n\left(\frac{\lambda-2}{\lambda}\right)^n.
            \label{Exf}
\end{aligned}
\end{equation}
By plugging Eqs.~\eqref{kappa} and ~\eqref{Exf} into Eq.~\eqref{GA2},  we obtain the recurrence relation
\begin{equation}
\begin{aligned}
            0 = & D^i_0-\frac{C^i_0 E^i_1}{D^i_1-}\frac{C^i_1 E^i_2}{C^i_2-}\frac{C^i_2 E^i_3}{D^i_3-}\ldots \,,
            \label{qEA}
\end{aligned}
\end{equation}
where $i=\kappa,\ \lambda$ and
\begin{equation}
    \begin{aligned}
 C^\kappa_n=&-2 (1 + n) (1 + m + n),\\
 D^\kappa_n=&-A_{lm}+2\tilde{a}^2\tilde{\epsilon}+2\tilde{a}(1-q)+(m + n)  (1 + m + n)\\
&-2\sqrt{-2\tilde{a}^2\tilde{\epsilon}}(1+m+2n),\\
E^\kappa_n=&-2\sqrt{-2\tilde{a}^2\tilde{\epsilon}}(m+n)+2\tilde{a}(q-1),\\
 C^\lambda_n=&(1 + n) (1 + m + n),\\
       D^\lambda_n=&-A_{lm}+2\tilde{a}^2\tilde{\epsilon}+2\tilde{a}(1+q)-(1+m)(1+2n)-2n^2\\
&+2\sqrt{-2\tilde{a}^2\tilde{\epsilon}}(1+m-2n)-\frac{2\tilde{a}(1+q)n}{\sqrt{-2\tilde{a}^2\tilde{\epsilon}}},\\
        E^\lambda_n=&\left(n+\frac{\tilde{a}(1+q)}{\sqrt{-2\tilde{a}^2\tilde{\epsilon}}}\right)\left(n+m+\frac{\tilde{a}(1+q)}{\sqrt{-2\tilde{a}^2\tilde{\epsilon}}}\right).
    \end{aligned}
\end{equation}

The roots of Eq.~\eqref{qEA} can be obtained starting from the limit of $\tilde{a}\ll1$, where we have $\tilde{\epsilon} \simeq \alpha'^2\mu/(2n^2)$ with $\alpha'=GM_1(1+q)\mu$. Then by stepping up $\tilde{a}$ in small increments, we can obtain the splitting structure of energy and angular eigenvalues for gravitational molecules at a certain orbital separation.

\subsection{Convergence test}

\begin{figure}
    \centering
    \includegraphics[width=0.45\linewidth]{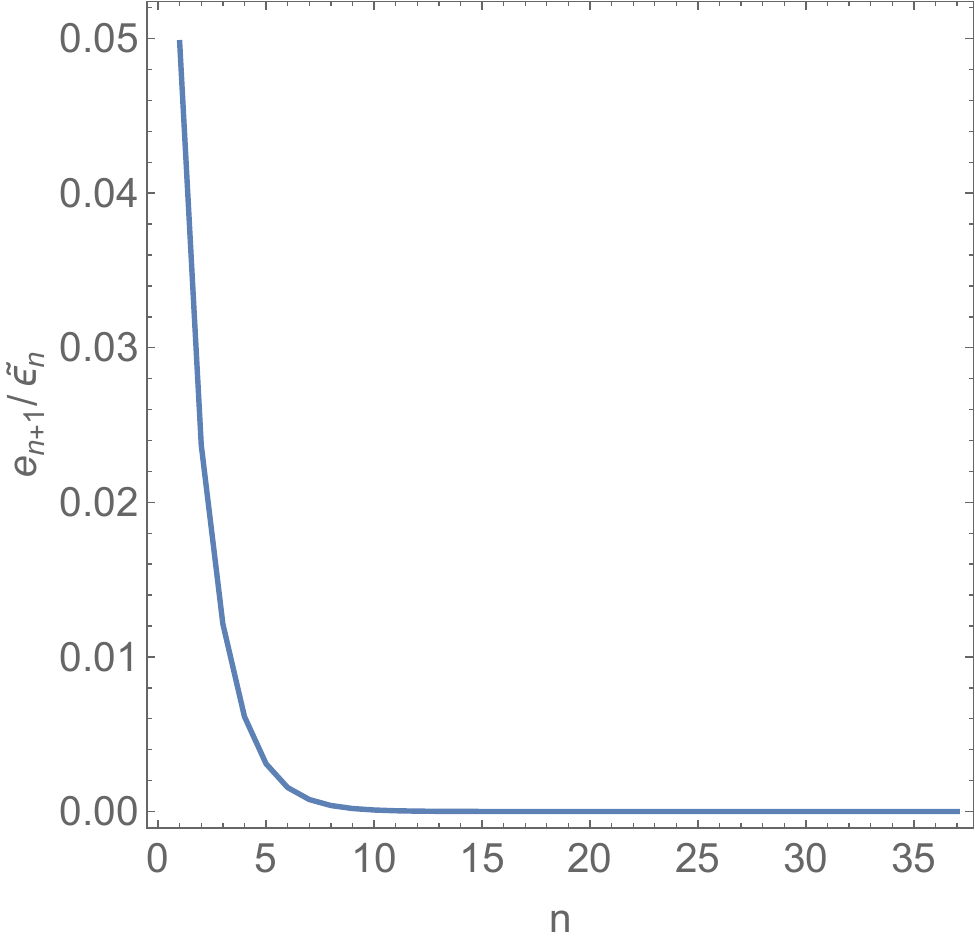}
    \includegraphics[width=0.45\linewidth]{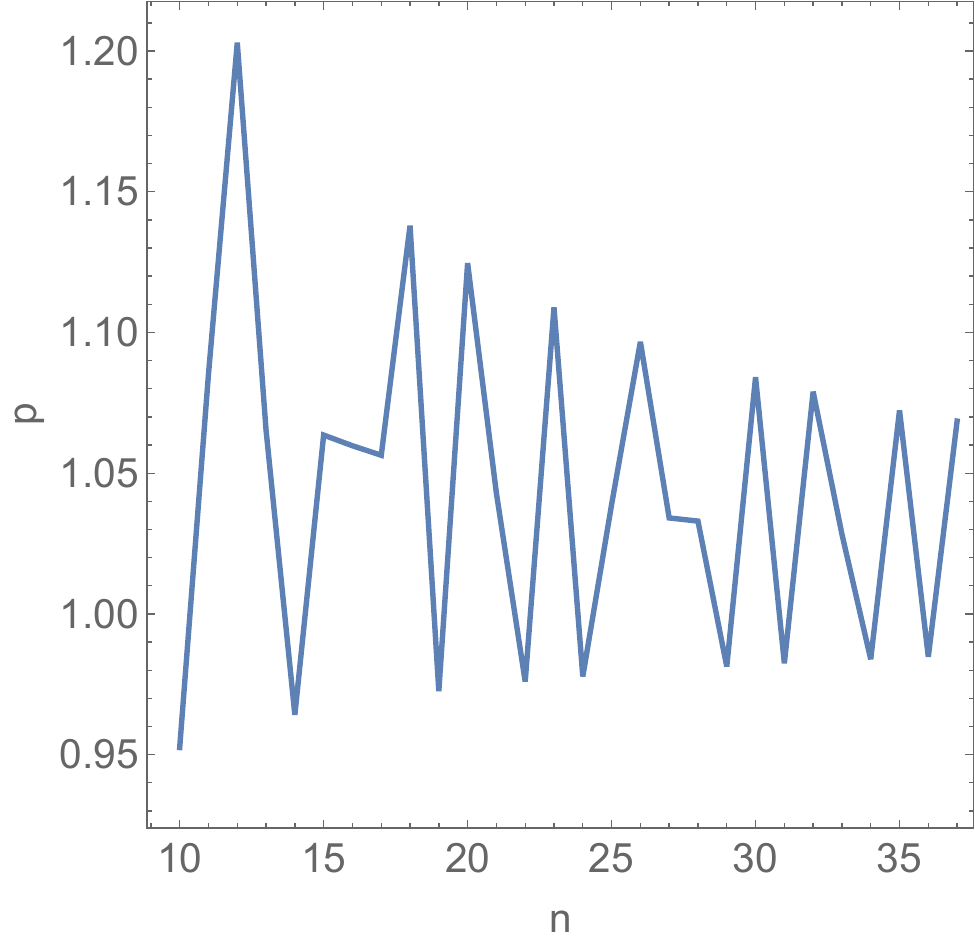}
    \caption{Convergence test of the shooting method used to solve the eigenenergy of $\bar\psi_{210}$ at the separation $\tilde{a}=0.1$. Both $e_{n+1}/\tilde{\epsilon}_n$ approaching zero and $p_n$ approaching $1$ indicate the convergency of the numerical computation.}
    \label{fig:shootE}
\end{figure}

\begin{figure}
    \centering
    \includegraphics[width=0.8\linewidth]{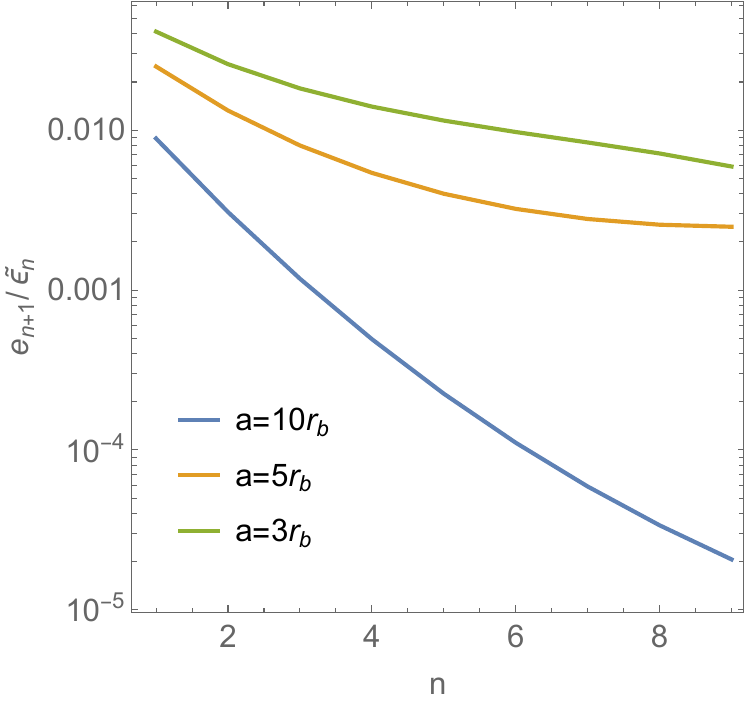}
    \caption{Convergence test of the continued fraction method used to solve the eigenenergy of $\bar\psi_{211}$ for $q=1$. The $e_{n+1}/\tilde{\epsilon}_n$ approaches zero indicating the convergency of the numerical computation, where colors of curves represent the eigenenergy at various orbital separations}
    \label{fig:shootyb}
\end{figure}

For both shooting and continued fraction methods, the convergence can be tested with
\begin{equation}
    e_n \equiv |\tilde\epsilon_n-\tilde\epsilon_{n-1}| \,.
    \label{eq:error}
\end{equation}  
For the shooting method, $n$ represents the iterations of fine-tuning the shooting parameters. For the continued fraction method, $n$ is the cutoff order of Eq.~(\ref{qEA}). When computing the eigenenergies, we demand $e_{n+1}/\tilde\epsilon_n\lesssim10^{-2}$ to guarantee the precision. In addition, we examined
\be
p (n) \equiv  \frac{\log |e_{n+1} - e_n|}{\log |e_n-e_{n-1}|}\,,
\ee
which approaches $1$ at large $n$, indicating the linear convergency of the results~\cite{stoer1980introduction}. For example, in Figs.~\ref{fig:shootE} and ~\ref{fig:shootyb}, we show the behavior of $e_n$ and $p_n$ for certain numerical results. The other results manifest a similar convergency behavior.


\section{GRAVITATIONAL ATOM IN PARABOLIC COORDINATE}
\label{app:para}

While the separable eigenfunctions of the gravitational atom are usually studied in spherical coordinates, they can also be obtained in parabolic coordinates~\cite{Castillo:2008} . The parabolic coordinates are defined by
\begin{equation}
    x=\sqrt{\rho\sigma}\cos{\chi},\quad y=\sqrt{\rho\sigma}\sin{\chi},\quad z=\frac{1}{2}(\rho-\sigma),
\end{equation}
and the time-independent Schrödinger equation in the coordinates can be written as
\begin{equation}
\begin{aligned}
    E\bar{\varphi}=&-\frac{2}{\mu(\rho+\sigma)}\left[\frac{\partial}{\partial\rho}\left(\rho\frac{\partial}{\partial\rho}\right)+\frac{\partial}{\partial\sigma}\left(\sigma\frac{\partial}{\partial\sigma}\right)\right]\bar{\varphi}\\
    &-\frac{1}{2\mu\rho\sigma}\frac{\partial^2}{\partial\chi^2}\bar{\varphi}-\frac{2\alpha}{\rho+\sigma}\bar{\varphi}.
    \label{eq:paraH}
    \end{aligned}
\end{equation}
By substituting the ansatz
\begin{equation}
    \bar{\varphi}=R_p(\rho)S_p(\sigma)e^{im\chi}
\end{equation}
into Eq.~\eqref{eq:paraH}, we can separate the variables and get
\begin{equation}
    \begin{aligned}
        \frac{d}{d\rho}\left(\rho\frac{dR_p}{d\rho}\right)+\left(\frac{E\mu}{2}\rho-\frac{m^2}{4\rho}+\beta_1\right)R_p&=0,\\
        \frac{d}{d\sigma}\left(\sigma\frac{dS_p}{d\sigma}\right)+\left(\frac{E\mu}{2}\sigma-\frac{m^2}{4\sigma}+\beta_2\right)S_p&=0,
        \label{eq:parasep}
    \end{aligned}
\end{equation}
where $\beta_1+\beta_2=\alpha\mu$. The solutions to Eq.~\eqref{eq:parasep} is known~\cite{landau2013quantum}, and the eigenstates are given by
\begin{equation}
    \begin{aligned}
        \bar{\varphi}_{n_s km}&=e^{im\chi}(\rho\sigma)^{|m|/2}L^{|m|}_{k}(\frac{\sigma}{n})L^{|m|}_{n_s-k-|m|-1}(\frac{\rho}{n})e^{-(\rho+\sigma)/(2n_s)},
        \label{wavfPara}
    \end{aligned}
\end{equation}
with $E=-\frac{\alpha^2\mu}{2n_s^2}$, where $L$ is the associated Laguerre polynomials. The relation between the eigenstates~\eqref{wavfPara} and the usual eigenstates obtained in the spherical coordinates, i.e., $\bar{\varphi}_{nlm}$,  has been investigated in Ref.~\cite{Castillo:2008}. Here we simply present the results
\begin{equation}
    \bar{\varphi}_{nlm}=\sum^j_{m_1,m_2=-j}(-1)^{j-m_2}\langle j,m_1,j,m_2|j,j,l,m\rangle\bar{\varphi}_{n_skm},
    \label{eq:sptopara}
\end{equation}
where 
\begin{equation}
    \begin{aligned}
        &n=2j+1,\quad m=m_1+m_2,\\
        &k=\left\{\begin{aligned}
            &m_2-m_1,\quad m_2-m_1>0\ ||\  m_1=m_2\neq0,\\
            &j,\quad m_1=m_2=0,\\
            &2j-|m_1+m_2|+m_1-m_2,\quad m_2-m_1>0,
        \end{aligned}\right.
        \label{corr}
    \end{aligned}
\end{equation}
and $\langle j,m_1,j,m_2|j,j,l,m\rangle$ denotes the Clebsch-Gordan coefficients. We refer readers to Ref.~\cite{Castillo:2008} for the derivation of the relation.

\section{RELATION BETWEEN THE ATOMIC EIGENSTATES AND THE MOLECULAR EIGENSTATES}
\label{sec:GAGM}

In this appendix, we demonstrate how to obtain the relation between the atomic eigenstates and the molecular eigenstates in the large separation limit with the prescription described in Sec.~\ref{sec:ceinit}. For illustration, we shall consider the two spin-aligned states $\varphi_{211}$, and derive Eq.~\eqref{eq:init} step by step. We shall refer to the spin directions of the two black holes as $\hat{\bf z}_1$ and $\hat{\bf z}_2$ and the direction of $\mathbf{r_2-r_1}$ as $\hat{\bf x}$. Since the two atomic states have the same eigenenergy, the involved molecular states should also have the same eigenenergy. Therefore, we can focus on the spatial part of the wave functions. 

At large orbital separation, the wave function is just the superposition of two atomic states
\begin{equation}
    \psi_0= \varphi_{1,211}+ \varphi_{2,211}\, .
\end{equation}
By changing to the corotating frame, it introduces an extra phase,
\begin{equation}
    \psi_0= e^{-i\Omega t }\left(\bar{\varphi}_{1,211}+ \bar{\varphi}_{2,211}\right).
\end{equation}
Note that $\bar{\varphi}_{1,211}$ and $\bar{\varphi}_{2,211}$ are eigenstates of angular momentum in $\hat{\bf z}_1$ and $\hat{\bf z}_2$, while the molecular eigenstates are eigenstates of angular momentum in $\hat{\bf x}$. We first decompose $\bar{\varphi}_{1,211}$ and $\bar{\varphi}_{2,211}$ into the atomic eigenstates of angular momentum in $\hat{\bf x}$,
\begin{equation}
   \psi_0= e^{-i\Omega_0 t}\sum_{i=1,2}\left(\frac{1}{2}\bar{\varphi}_{i,21-1}^{\hat{x}}-\frac{1}{\sqrt{2}}\bar{\varphi}_{i,210}^{\hat{x}}+\frac{1}{2}\bar{\varphi}_{i,211}^{\hat{x}}\right).
    \label{eq:ZX}
\end{equation}
Next, we decompose each atomic eigenstate into atomic eigenstates in parabolic coordinates using Eq.~\eqref{eq:sptopara},
\begin{equation}
    \psi_0= e^{-i\Omega_0 t}\sum_{i=1,2}\left(-\frac{1}{2}\bar{\varphi}^{(p)}_{i,20-1}+\frac{1}{2}\bar{\varphi}^{(p)}_{i,200}-\frac{1}{2}\bar{\varphi}^{(p)}_{i,210}+\frac{1}{2}\bar{\varphi}^{(p)}_{i,201}\right).
    \label{eq:paraGA}
\end{equation}
Finally, using the relation~\eqref{qrelation} derived from node conservation~\cite{BATES196813}, we can express
\begin{equation}
 \bar{\varphi}^{(p)}_{1,{\rm n_s k m}} + \bar{\varphi}^{(p)}_{2,{\rm n_s k m}} = \sqrt{2}\bar{\psi}_{n \ell m} \,.
\end{equation}
In particular, we have
\begin{equation}
\begin{aligned}
   & \bar{\varphi}^{(p)}_{1,{\rm 20-1}} + \bar{\varphi}^{(p)}_{2,{\rm 20-1}} = \sqrt{2}\bar{\psi}_{21-1}\,,\quad
     &\bar{\varphi}^{(p)}_{1,{\rm 200}} + \bar{\varphi}^{(p)}_{2,{\rm 200}} = \sqrt{2}\bar{\psi}_{310}\,, \\
    & \bar{\varphi}^{(p)}_{1,{\rm 210}} + \bar{\varphi}^{(p)}_{2,{\rm 210}} = \sqrt{2}\bar{\psi}_{430}\,,\quad
     &\bar{\varphi}^{(p)}_{1,{\rm 201}} + \bar{\varphi}^{(p)}_{2,{\rm 201}} = \sqrt{2}\bar{\psi}_{211}\, ,
     \label{eq:PATOGM}
\end{aligned}
\end{equation}
substituting which into Eq.~\eqref{eq:paraGA} leads to
\begin{equation}
    \psi_0= \frac{e^{-i \Omega_0 t}}{\sqrt{2}}\left(\bar{\psi}_{211}-\bar{\psi}_{21-1}+\bar{\psi}_{310}-\bar{\psi}_{430}\right) \, ,
\end{equation}
and hence Eq.~\eqref{eq:init}. Equation~\eqref{eq:init2} can be obtained in the same way. One can further express a single atomic state, e.g., $\varphi_{1,211}$ in terms of molecular states, and hence construct the initial condition even if the clouds surrounding the two black holes are in different atomic eigenstates or with nonparallel spin alignment. Nevertheless, this relation is valid only for equal-mass binaries.

\newpage

\bibliography{references}

\end{document}